\documentstyle[psfig]{article}
\def\m@thcombine#1#2{%
  \setbox0=\hbox{$#1$}
  \setbox1=\hbox{$#2$}
  \ifdim\wd0>\wd1
    \setbox0=\hbox to\wd1{\hss\box0\hss}
  \else
    \setbox1=\hbox to\wd0{\hss\box1\hss}
  \fi
  \mathop{\vcenter{
    \offinterlineskip\box0\box1}}}
\def\lesim{\m@thcombine<\sim}
\def\gesim{\m@thcombine>\sim}
\def\lessgtr{\m@thcombine<>}
\def\gtrless{\m@thcombine><}
\newcommand{\bra}[1]{\left\langle #1 \right|}
\newcommand{\ket}[1]{\left| #1 \right\rangle}
\newcommand{\omegaI}{\omega_I}
\newcommand{\omegaII}{\omega_{I+2}}
\newcommand{\eps}{\epsilon}
\newcommand{\beq}{\begin{equation}}
\newcommand{\beqa}{\begin{eqnarray}}
\newcommand{\eeq}{\end{equation}}
\newcommand{\eeqa}{\end{eqnarray}}
\newcommand{\Yb}{${}^{168}$Yb\ }

\begin{document}

\LARGE

\begin{center}
{\bf Shell Model for Warm Rotating Nuclei}
\end{center}

\large

\vspace{10mm}

\begin{center}
 M. Matsuo, T. D\o ssing$^{a}$, E. Vigezzi$^{b}$,
R.A. Broglia$^{a,b}$, and K. Yoshida$^{c}$
\end{center}

\vspace{10mm}

\large

\begin{center}
{\it
Yukawa Institute for Theoretical Physics, Kyoto University,
Kyoto 606-01, Japan \\
${}^{a}$ Niels Bohr Institute, University of Copenhagen,
Copenhagen \O, DK2100, Denmark \\
${}^{b}$ INFN sez. Milano, and Dept. of Physics, University of Milan, Milan,
Italy \\
${}^{c}$ Department of Physics, Kyoto University,
Kyoto 606-01, Japan
}
\end{center}

\vspace{10mm}

\normalsize

\begin{abstract}
In order to provide a microscopic description of levels and E2 transitions
in rapidly rotating nuclei with internal excitation energy up to
a few MeV, use is made of a shell model which combines the
cranked Nilsson mean-field and the residual surface delta  two-body
force. The damping of collective rotational motion
is investigated in the case of a typical
rare-earth nucleus, namely \Yb.
It is found that rotational damping sets in
at around 0.8 MeV above the yrast line,  and  the levels
which form  rotational band structures are thus limited.
We predict at a given rotational frequency existence
of about 30 rotational bands of various lengths,
in overall 
agreement with the experimental findings.
The onset of the rotational damping proceeds quite gradually
as a function of the internal excitation energy. The transition
region extends up to around 2 MeV above yrast and it is characterized
by the presence of scars of discrete  rotational bands which
extend over few spin
values and stand out among the damped
transitions, and by  a two-component profile in the $E_\gamma -E_\gamma$
correlation. The important role played by the high-multipole components of the
two-body residual interaction is emphasized.

\vspace{5mm}


\end{abstract}

\section{Introduction}\label{sec:intro}

A heavy ion fusion reaction produces a hot and rapidly rotating compound
nucleus. After emission of neutrons has cooled down the system, the
compound nucleus still keeps high
angular momenta (tens of $\hbar$) and
moderate heat energy, that is a moderate excitation energy (a few MeV) once
the energy of rotational motion has been subtracted.
It then emits
a multitude of $\gamma$-rays, decreasing its spin and heat energy
gradually, reaching finally the ground state.

For nuclei with mass number $A \sim$ 170 whose  ground state has a stable
elongated shape ($\beta \sim 0.3$), up to around 20 different
rotational bands
have been observed
in discrete gamma ray spectroscopy studies.
The observed
rotational bands lie
in the lowest excitation energy region near the yrast line, representing
the  ``cold'' part of the whole of nuclear levels.
In  the excited region,
say at 1 to 2 MeV above the yrast line,
the level density at a given spin is as large as $10^2$ to
$10^3 $ MeV$^{-1}$ in  normally deformed rare-earth nuclei.
The equivalent  level spacing of 10 to 1 keV is rather small
in comparison to
the typical size ($ \sim 10 $keV) of  matrix elements
of the residual two-body nuclear force.
The residual interaction then becomes effective
to cause  mixing of many-particle many-hole (or many-quasiparticle)
configurations in the rotating deformed mean-field potential.
Since  different configurations respond differently to the
Coriolis force
due to different single-particle alignments,
the configuration mixing results in a dispersion of the rotational
frequency within each energy eigenstate, implying a damping of 
the collective rotational motion
\cite{Leander,Lauritzen}.
The gamma-rays
which are emitted from the ``warm'' region (from about 1 MeV
 to about a  few MeV above yrast, corresponding to temperature
of a few hundred keV ) cannot be distinguished as  discrete
peaks and form  a quasi-continuum in the spectra.
Rotational damping has been studied  experimentally
through the analysis of the quasi-continuum spectra
\cite{Bacelar,Draper,Stephens}.
Recently,
new experimental techniques have been devised to deal with
the multi-dimensional quasi-continuum gamma-spectra
obtained by double or triple coincidence gamma-ray experiments,
making it possible to study in detail various aspects of
the collective rotational motion in the warm region
\cite{FAM-168Yb,FAM-PR,Rev,RPM,Erice,Leoni}.
In particular,  the fluctuation analysis method 
\cite{FAM-168Yb,FAM-PR}
has revealed that
the number of rotational bands existing in a rare-earth
nucleus is only around 30 at a given rotational
frequency, thus confirming the occurrence of the
rotational damping.

Early theoretical studies of the rotational damping dealt with 
the E2 strength function associated with the damped rotational motion,
which is obtained by assuming that the configuration mixing
is described by the general statistical
theory of random matrices
\cite{Leander,Lauritzen,Guhr,Egido}.
In this paper we present 
a  microscopic shell model able 
to describe individual nuclear levels and E2 transitions
in the warm region, 
in order 
to study in more detail the transition 
from the regime of discrete rotational bands into the
regime of damped rotational motion.
Our description is based on a mean field calculated by the cranking 
model, which has been quite successful in the description of the 
low-lying portion of the nuclear spectrum
(cf. e.g. \cite{Aberg-Flocard-Nazarewicz,Szymanski,de-Voigt,
Bengtsson-Frauendorf,Bengtsson-Ragnarsson}).
We then add a
two-body effective interaction, mixing the mean-field configurations.
The first attempt along this line was  made
in Ref.\cite{Aberg}, where  a  rather
schematic residual interaction was employed in the calculations.
We instead adopt 
somewhat more "realistic"
effective nuclear force, i.e.,
the surface delta force \cite{Mozkowski}.
We found in previous studies\cite{Matsuo93,MatsuoNPA} that the
high-multipole components, which are present in this force, 
are  essential to produce the rotational damping.
In the present formulation, the emphasis is
put not only on 
the residual two-body force but also on the construction
of the diabatic single-particle basis, which enables us to
describe long sequences of the rotational E2 transitions.
To be noted is that both the
particle-rotor model \cite{Kruppa}
and the interacting boson model \cite{ibm} have been used to attack
problems  similar to the one we are dealing with in the present paper.
Both of these models are more complete than the present
one with respect to the angular momentum coupling in wave functions
and E2 transition matrix elements. However, they
are severely restricted by taking into account only a
small part of many-particle many-hole excitations
(single-$j$ orbits and $s$,$d$-bosons, respectively).
The present approach includes all the
excitations in the single-particle orbits in deformed nuclei within a
given excitation energy.

After describing the formulation  in Sect.\ref{sec:form},
we present numerical results for \Yb in Sect.\ref{sec:results}.
We investigate in detail
the excited  states
in the warm region of the nuclear spectrum
with very high spins as well as associated E2
transitions, which characterize the collective rotational motion.
In particular,  we look into the properties of rotational damping and
compare them with available experimental data.
In Sect.\ref{sec:resint}, we study in detail 
high multipole components
of the two-body effective force and compare
the surface delta interaction with the
pairing plus quadrupole-quadrupole force.

\section{Formulation}\label{sec:form}

\subsection{Single-particle basis}

We consider deformed nuclei with stable prolate shape.
In such a nucleus,
rotational bands observed by the discrete gamma-spectroscopy
are often interpreted in terms of the single-particle excitations
in  cranked mean-field models where the collective rotational excitation
is described in a semi-classical way while the intrinsic
particle excitations are described quantum mechanically.
For example, the cranked Nilsson model is widely used
and gives a realistic description of the single-particle
excitation modes in the  high spin yrast region
\cite{Bengtsson-Frauendorf,Bengtsson-Ragnarsson}.
Let us assume that the  potential  surface
at a given spin
displays a deep minimum  in the space of the deformation parameters
for a prolate
deformation.
In such a case,
it should be possible to have intrinsic excitations
of many particles and holes (or many quasiparticles)
on top of the stable  deformed mean field state.
The excited states above the yrast line are to be
built out of these excited configurations.

In keeping with
this picture, we formulate a shell model in which
the single-particle basis is
represented by the cranked Nilsson model and the
many-particle many-hole ($n$p-$n$h) excitations associated with
the single-particle potential are taken into account to represent
the intrinsic excitations in rotating nuclei.
The collective E2 transitions are assumed to
keep the intrinsic structure, that is, each of
the $n$p-$n$h
configurations forms a rotational band.
Since the level density of the $n$p-$n$h states is high, the residual
two-body force
causes mixing among the $n$p-$n$h excitations, that is,
mixing among the rotational bands.
The configuration mixing can be taken into account
explicitly by solving a shell model Hamiltonian which combines
the cranked Nilsson mean-field and the residual two-body force

In the following, we describe the details of the model.
Let us start with the single-particle basis.
The single-particle Hamiltonian in the present theory
is given by the cranked Nilsson model
\beq
	h_{crank} = h_{Nilsson} - \omega j_x,
\label{nilham}
\eeq
in which
the protons and neutrons move in a
deformed Nilsson potential which rotates  around the
axis with the
largest moment of inertia
(taken conventionally as  the $x$ axis), with
uniform rotational frequency $\omega$.
We shall not include the pairing potential
in the mean-field Hamiltonian.
Note,  however, that pairing correlations may be partly
taken into account
through the configuration mixing
caused by the residual interaction.
Moreover,
we shall be mainly interested in
levels at very high spins $I \gesim 30$, where the pairing
gap is expected either to be  small or to vanish,
due to the Coriolis anti-
pairing effect \cite{Garrett,Shimizu,Shimizu-Oak}.
Inclusion of the pairing potential in $h_{crank}$
may improve the description of near-yrast rotational bands at low spins.
However,
use of the quasiparticle scheme may lead to incorrect description
of highly excited states since many-quasiparticle configurations accompany
spurious components
related to  the nucleon number violation.
We use  the Nilsson potential  \cite{Nilsson}
with a single-stretched $l^2$ term and an $ls$ force
whose parameters are given in
\cite{Bengtsson-Ragnarsson}. The quadrupole and hexadecapole
deformation parameters of the Nilsson
potential $(\eps, \eps_4)$  should in principle  be determined
selfconsistently by minimizing the potential energy surface.
The numerical calculations presented
in the present paper  were carried out with the deformation parameters 
$(\eps, \eps_4) = (0.255, 0.014)$ for \Yb, according to
Ref.\cite{Def-parm}. The single-particle basis
keeps the signature and parity quantum numbers.

\begin{figure}
\psfig{figure=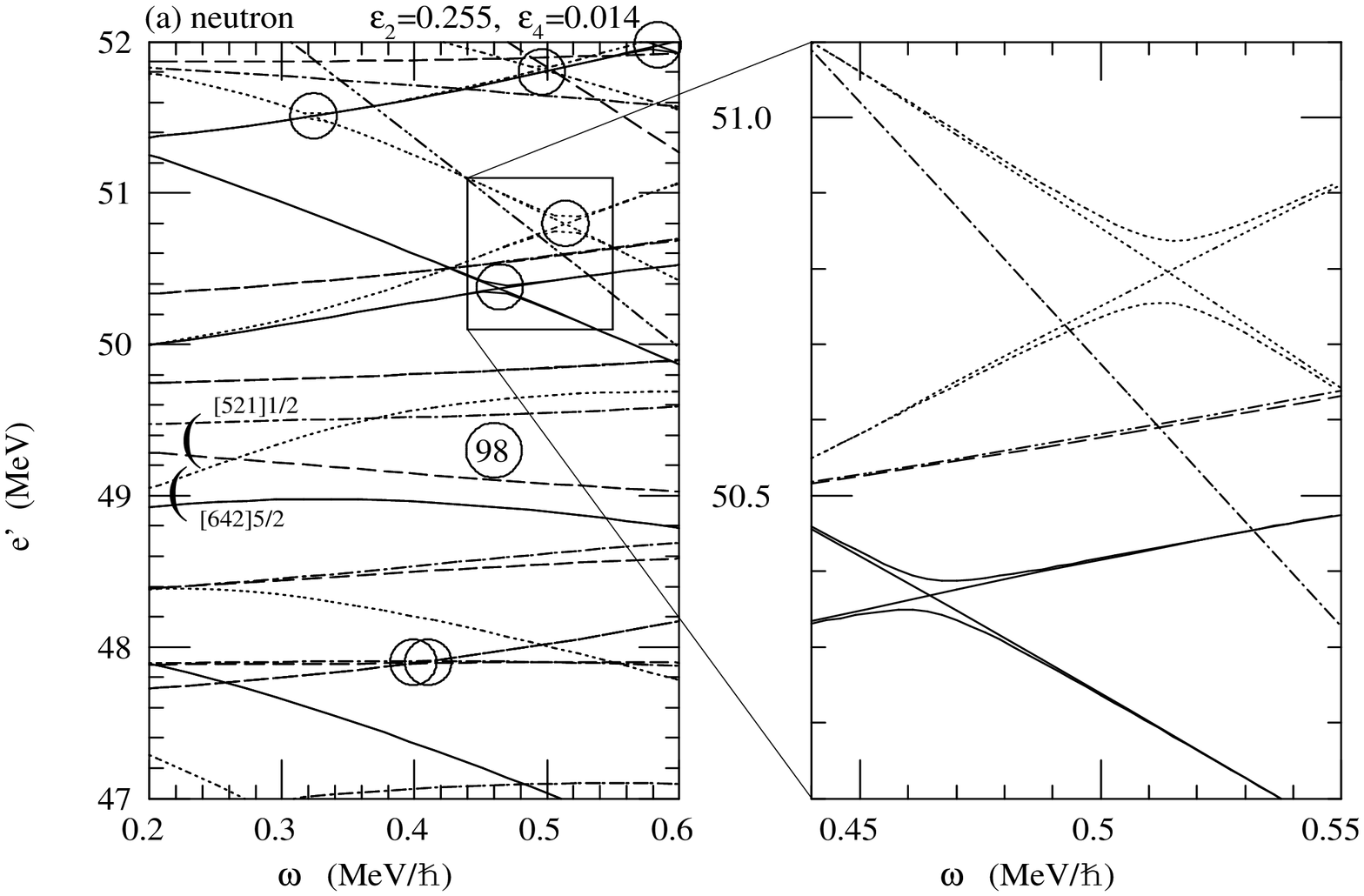,height=9cm,angle=-90}
\vskip 5mm
\psfig{figure=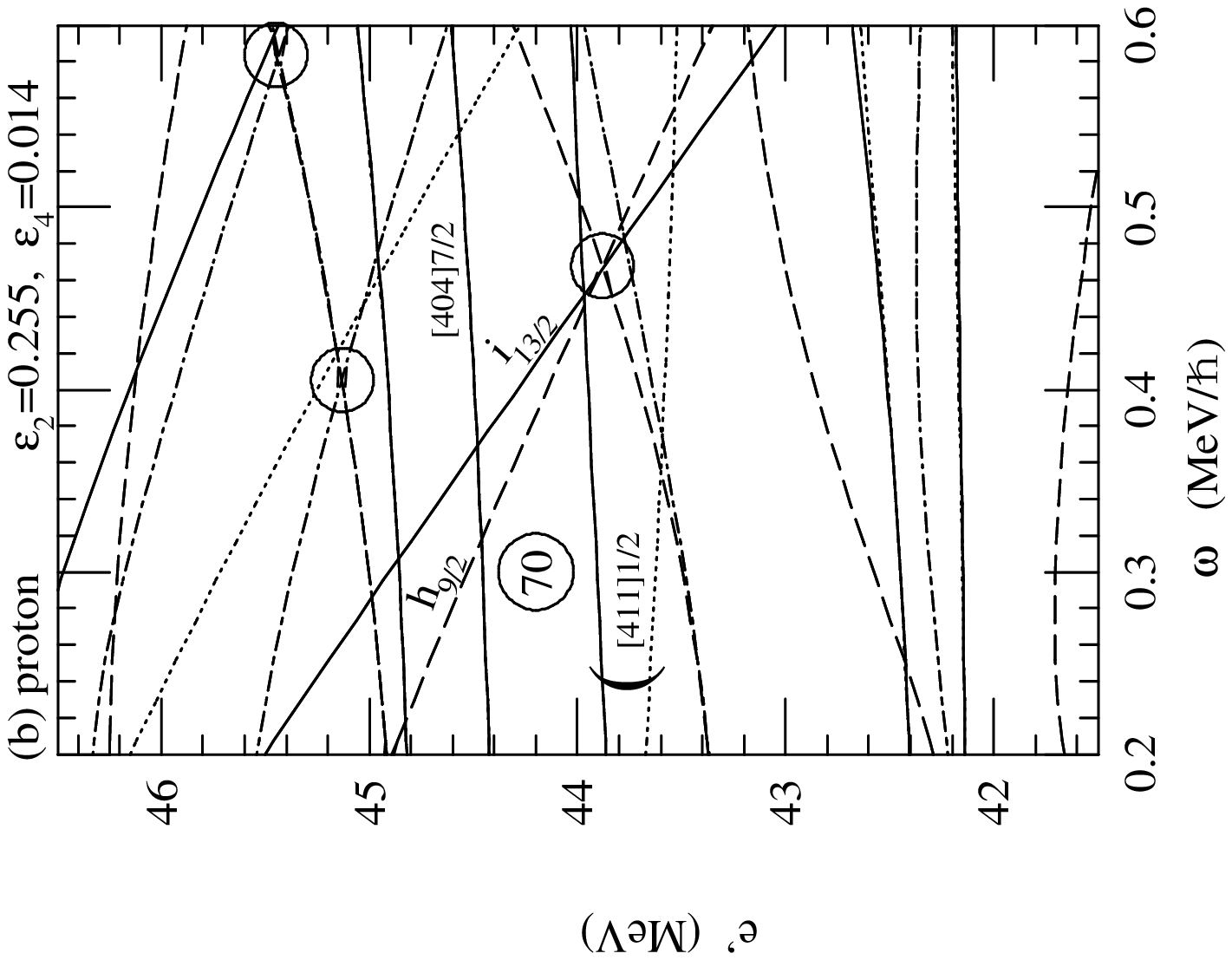,height=9cm,angle=-90}
\caption{
The cranked Nilsson single-particle routhian
spectra for (a) neutrons  and (b) protons 
with deformation parameter $(\eps,\eps
_4)=(0.255,0.014)$.
The different kinds of curves denote different
parity and signature; $(\pi,\alpha)=(+,1/2),(+,-1/2),(-,1/2)$
and $(-,-1/2)$ for solid, dotted, dashed and dot-dashed
curves, respectively.
The circles indicate the positions where
the pair-wise repulsions take place. Both adiabatic and diabatic
basis are plotted while they mostly overlay with each other.
In the right panel, we
show a magnified portion of the neutron routhian spectrum
so that difference between the two basis becomes visible.
\label{fig1}}
\end{figure}

The eigenstates of the cranked Nilsson Hamiltonian, Eq.(\ref{nilham}),
define the adiabatic single-particle orbits. The eigenrouthians
and the single-particle wave functions vary with
the rotational frequency $\omega$. From time to time
pairs of orbits cross each other as a function of $\omega$
(See fig.\ref{fig1}).
If we magnify the crossing among the orbits with the same quantum
numbers, it is seen that the two orbits ``repel'' each other.
The nature
of the two orbits interchanges across the crossing. This
causes an abrupt change in the single-particle  orbits
as a function of the rotational frequency. This abrupt
change makes it difficult to define the rotational bands
on the basis of the single-particle orbits given adiabatically.
To remedy this
problem and guarantee the smooth change as a function of
$\omega$, a {\it diabatic basis} for the single-particle orbits 
\cite{Bengtsson} is
constructed in place of the adiabatic eigen solutions.
The method of constructing the diabatic basis is described in the
Appendix A. The resulting
diabatic routhian spectrum
$\{e'_i(\omega)\}$ is also
shown in Fig.\ref{fig1}. In the following, we adopt
the diabatic single-particle basis
with routhians $\{e'_i(\omega)\}$, single-particle wave functions
$\{\psi_i(\omega)\}$ and the associated
angular momentum expectation value  $\{ j_{x,i}(\omega)\}$.

\subsection{Unperturbed rotational bands}\label{sec:base}

Once the single-particle basis is defined, the
shell model basis
of the many-body system is obtained by filling
$N$ neutrons and $Z$ protons in the
diabatic cranked Nilsson  orbits.
Based on the configuration where all the orbits up to the
Fermi surface are occupied, one can generate
many-particle many-hole ($n$p-$n$h) excitations,
which form shell model basis states
for the excited states above the yrast line.
The basis configurations, labeled by $\mu$, at spin $I$ 
\footnote{In this paper, the spin is measured in units of
$\hbar$}
are given by
\beq
\ket{\mu (I)} = \prod_{{\rm occupied}\ i \ {\rm in} \ \mu} a_i^{\dag} \ket{0}
\label{mu}
\eeq
where $a_i^{\dag}$ represents the  creation operator for the
diabatic cranked
Nilsson single-particle wave function $\psi_i(\omegaI)$ occupied
in this configuration. The rotational frequency $\omega_I$
corresponding to the spin value $I$ is calculated by
imposing the condition
\beq
\langle J_x \rangle (\omega_I) = I
\label{jx}
\eeq
on the average value of the projection 
$\langle J_x \rangle $ of the angular momentum on the
rotation axis. Here this quantity is calculated by using 
thermal averaging as
\beq
\langle J_x \rangle  = \sum_i j_{x,i}(\omega) f_i(\omega) \ \ ,
\eeq
where $j_{x,i}(\omega)$ is the expectation value of $j_x$ of an orbit
$i$, and  $f_i(\omega) = \{ 1 + \exp{(e'_i(\omega)- \lambda)/T}\}$
is the thermal occupation probability. The
temperature parameter is chosen as $T=0.4$ MeV so that it
corresponds to thermal excitation energy $U \sim 2 $ MeV relevant
to the excited states under discussion.

\begin{figure}
\psfig{figure=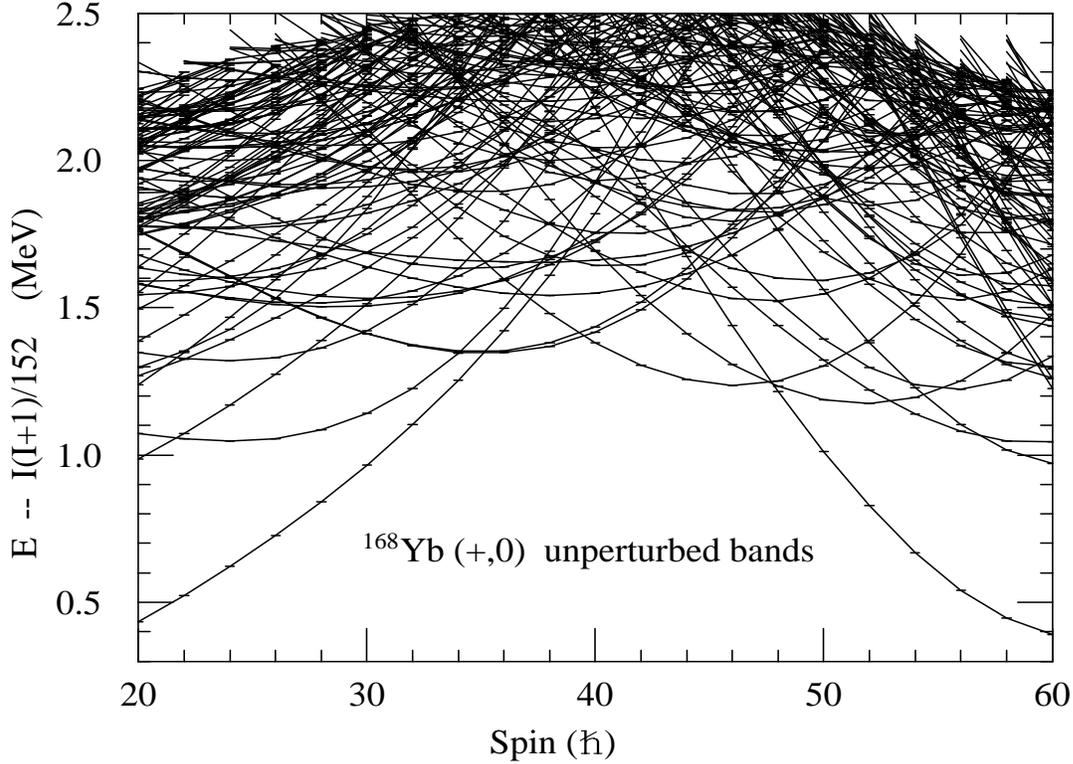,width=14cm,angle=-90}
\caption{
The unperturbed rotational bands in \Yb with
$(\pi,\alpha)=(+,0)$. The lowest one hundred levels
are plotted for each spin with small horizontal
bars. A reference energy
$I(I+1)/2J$ with $J=76$ MeV${}^{-1}$
is subtracted.
\label{fig2}}
\end{figure}

The energy of
the basis states in the laboratory frame 
is  given by the Strutinsky renormalization
method according to the prescription commonly used
in the cranked Nilsson model
\cite{Andersson,Neergard,Bengtsson-Ragnarsson}.
Namely, the energy
$E_\mu(I)$ of the basis state $\mu$ at spin $I$ is
given by
\beq
E_{\mu}(I) = E_{\mu}^{Nils}(I) - E^{smooth}(I) + E^{LD}(I)
\label{Str}
\eeq
Here the shell correction energy $E_{\mu}^{Nils}(I) - E^{smooth}(I)$
is calculated microscopically with use of the diabatic
single-particle basis. The bare  energy
$E_{\mu}^{Nils}(I)$ is defined by
$E_\mu^{Nils}(I)=E'_\mu(\omega) +\omega J_{x,\mu}(\omega) $ with
a constraint  $J_{x,\mu}(\omega) = I$ on the rotational
frequency $\omega$. Here $E'_\mu(\omega)$ and
$J_{x,\mu}(\omega)$ are the total routhian
and the expectation value of the angular momentum $J_x$, respectively.
Utilizing the diabatic single-particle routhian basis, 
$E'_\mu(\omega)$  and $J_{x,\mu}(\omega)$ change smoothly as a function
of the rotational frequency, and it is possible to approximate
the bare energy $E_{\mu}^{Nils}(I)$ in the laboratory frame as
\beqa
E_{\mu}^{Nils}(I) & = & E'_{\mu}(\omegaI) + \omegaI I +
{(I - J_{x,\mu}(\omegaI))^2 \over\ 2 J^{(2)}_{\mu}} , \\
\label{Emu}
E'_\mu(\omegaI) & = &\sum_{{\rm occupied}\ i \ {\rm in }\ \mu}
e'_i(\omegaI), \\
J_{x,\mu}(\omegaI) & = & \sum_{{\rm occupied}\ i \ {\rm in }\ \mu}
j_{x,i}(\omegaI), \\
J^{(2)}_{\mu} & = & \left.{ dJ_{x,\mu} \over d\omega
}\right|_{\omegaI} \ \ ,
\eeqa
by means of the second order extrapolation from the reference
rotational frequency $\omegaI$ \cite{Aberg,Level-density}.
The deviation $\left | J_{x,\mu}(\omega_I) - I \right |$ 
in the angular momentum expectation value is less than 5
at spin $I=50$ for most configurations in the present calculation.
The Strutinsky smoothed energy is given by
\beq
E^{\rm smooth}(I)=\sum_i e'_i(\tilde{\omega}) \tilde{n}_i+\tilde{\omega}I ,
\eeq
and
\beq
\tilde{J}_x(\tilde{\omega})=\sum_i j_{xi}(\tilde{\omega})\tilde{n}_i=I,
\eeq
with the smoothed occupation number 
$\tilde{n}_i$ \cite{Ring}.
The rigid-body rotational energy $E^{LD}=I(I+1)/2J_{rigid}$ is
calculated for the given shape of the potential
($J_{rigid} = 80.3$ MeV${}^{-1}$ for \Yb).
Since the single-particle
orbits have good parity $\pi$ and signature $\alpha$, the
shell model basis states
also conserve the same quantum numbers.

The energies $\{ E_\mu(I)\}$
and the many-body wave functions $\{ \ket{\mu(I)}\}$ define
the shell model basis at a given spin $I$ and parity $\pi$.
It is assumed that to each $n$p-$n$h configuration there
corresponds a rotational band. It is possible to make this 
correspondence uniquely and to follow the evolution of
the band as a function of $I$ in a continuous way
because the basis states depend smoothly on the rotational
frequency due to the diabatic construction of
the single-particle basis.
Such rotational bands 
are shown in Fig.\ref{fig2}
for \Yb.

\subsection{Residual two-body interaction and shell model diagonalization}

Since the mean-field potential is represented by
the Nilsson deformed potential, it is the residual part of
the two-body nuclear effective force
that is to be taken into account as the shell model
two-body interaction. In order to separate the residual part,
we utilize the fact that, given a
shell model configuration (a determinantal many-body state),
any two-body force is  decomposed
unambiguously into the mean-field  and the residual parts.
The residual two-body interaction $V_{\rm res}$ associated with the
reference shell model configuration, denoted
by $\mu_{\rm ref}$, is given by subtracting the mean-field
part $V_{\rm mf}$ from the two-body interaction;
\beq
V_{\rm res} = V_{\rm 2body}  - V_{\rm mf} 
 + \langle V_{\rm 2body} \rangle \ \ .
\eeq
Here the two-body interaction $V_{\rm 2body}$ is expressed as 
\beq
V_{\rm 2body} = \frac{1}{4} \sum_{ijkl}v_{ijkl} a_i^{\dag} a_j^{\dag} a_l a_k
\eeq
with $v_{ijkl}=\bra{\psi_i(1)\psi_j(2)-\psi_j(1)\psi_i(2)}
v(1,2)\ket{\psi_k(1)\psi_l(2)}$
being the anti-symmetrized  matrix elements of
the two-body force
for the single-particle orbits $i,j,k,$ and $l$. 
The mean-field part $V_{\rm mf}$ and a constant expectation
value $\langle V_{\rm 2body} \rangle$ is given by
\beqa
V_{\rm mf}    & = &\sum_{ij} v^{\rm mf}_{ij} a_i^{\dag} a_j\ \  , \ \ \
v^{\rm mf}_{ij} = \sum_{{\rm occupied}\ k \ {\rm in} \ \mu_{ref}}
v_{ikjk} \ \ , \\
\langle V_{\rm 2body} \rangle    &  = & {1 \over 2} 
\sum_{{\rm occupied}\ jk \ {\rm in} \ \mu_{ref}}
v_{jkjk}\ \ .
\eeqa
In this way, we could construct the residual interaction so that
it has no diagonal matrix elements
for the reference configuration, i.e.
$\langle \mu_{\rm ref}| V_{\rm res} |\mu_{\rm ref}\rangle = 0$.
In the case of a even-even nucleus, we
select as the reference $\mu_{\rm ref}$ the configuration
in which 
the single-particle routhian orbitals are filled
up to the Fermi level
at low rotational
frequency. We then take the diabatic continuation of
the same configuration to extend to  higher
rotational frequencies.
The residual interaction
thus determined is essentially independent on
spin or rotational frequency.

As the effective two-body effective force,
we adopt the surface delta interaction (SDI) \cite{Mozkowski}
with standard strength $V_0 = 27.5/A$
\cite{Faessler}(See the Appendix B for
details).
The SDI has  been
used widely in the shell model descriptions of
nuclei in a wide mass range from light nuclei to deformed
rare-earth.
The delta force acting only at the nuclear surface  displays
essential features needed for describing
low-lying excitations. Note that it contains the
the pairing force in the particle-particle channel as well as the
multipole-multipole forces in the particle-hole channel.
In a later section, we will investigate this two-body force
by changing its force strength as well as by comparing it with other 
effective forces.


Given the residual two-body interaction, the shell model
Hamiltonian for the basis states $\{\ket{\mu(I)}\}$ at spin $I$
is constructed as
\beq
H(I)_{\mu\mu'} = E_{\mu}(I) \delta_{\mu\mu'} +
V(I)_{\mu\mu'}
\eeq
\noindent
where $ V(I)_{\mu\mu'} =\langle\mu(I) |V_{\rm res}|\mu'(I)\rangle$
are the matrix elements  of
the residual two-body interaction.

The
energy eigenstates of the
shell model Hamiltonian are obtained by numerical diagonalization.
The resultant eigenstates are admixture of the shell model basis states,
that is,
the unperturbed rotational bands $\mu's$,
\beq
\ket{\alpha(I)} = \sum_\mu X^{\alpha}_{\mu}(I) \ket{\mu (I)}  \ \ .
\eeq
In carrying out the diagonalization,
truncation of the basis states is necessary.
For that purpose 
we first truncate the single-particle routhian orbitals so that
only those located within $\delta E_{max}$  above and below the Fermi
surfaces are treated as active orbits.
In the numerical calculations presented below, we choose
 $\delta E_{max}= 3$ MeV, which corresponds to  20-30 orbitals
for protons and for neutrons. 
We then consider all the $n$p-$n$h configurations
whose routhian excitation energy
$\delta E' = \sum_{particles} e'_i - \sum_{holes} e'_j$
lies within the interval $\delta E_{max}$. 
This generates about 2000-3000
$n$p-$n$h configurations for each $I^\pi$, most of which
have 1p1h to 4p4h character while
there is only a little contribution of 5p5h under the
energy cutoff. The dimension  of the basis is still too large
to carry out systematic numerical diagonalization. To avoid this difficulty,
we calculate the energy $\{E_\mu(I)\}$
of the basis states $\{\ket{\mu(I)}\}$ as well as the diagonal contribution
to the basis energy of the residual interaction
 $ V(I)_{\mu\mu}$ for each configuration.
The basis configurations are sorted up according to the energy
$E_\mu(I)+ V(I)_{\mu\mu}$ containing
the diagonal residual interaction, and the lowest 10$^3$ basis states
are included for the shell model diagonalization. The diagonalization
is done separately for each $I^\pi$.

\vskip 5mm
\subsection{ E2 transition strengths}

Using the assumption
$I \simeq I_x$, the electromagnetic
transition matrix elements can be calculated within the cranking
model \cite{Hamamoto-Sagawa}.
In order to calculate the stretched E2 transitions,
we  neglect minor contributions of
the non-collective E2  transitions so that only the
E2 matrix elements connecting the same
intrinsic configuration are taken into account.
Assuming that
the nuclear shape is
stable and independent of the configuration,
the E2 operator then becomes proportional to a
constant quadrupole moment $Q_o$;
\beqa
M_{\mu I,\mu' I-2} &  =  &
\bra{\mu (I)} M(E2,\lambda=2,\mu_x=2)\ket{\mu'(I-2)} \\
           &  = & \sqrt{15\over 128\pi} Q_o \delta_{\mu\mu'}  \ \ ,
\eeqa
where $\delta_{\mu\mu'} = 1$ if the configuration  $\ket{\mu(I)}$
is the diabatic continuation of
$\ket{\mu'(I-2)}$, and $\delta_{\mu\mu'}=0$ for other configurations.

The stretched E2 transition probability among the
energy eigen states $\alpha$ at $I$ and $\alpha'$ at $I-2$
is calculated as
\beq
B(E2,\alpha I \rightarrow \alpha' I-2) = {15 \over 128\pi}Q_o^2
M_{\alpha I,\alpha' I-2}^2
\eeq
with amplitude
\beq
M_{\alpha I,\alpha' I-2} = \sum_{\mu}X^{\alpha}_{\mu}(I)
X^{\alpha'}_{\mu}(I-2) \ \ .
\eeq
Hereafter, we often use the normalized E2 strength
\beq
S_{\alpha I,\alpha' I-2} \equiv
M_{\alpha I,\alpha' I-2}^2 \ \ ,
\eeq
which satisfies
$\sum_{\alpha'} S_{\alpha I,\alpha' I-2} = 1$, for E2
decays from the level $\alpha$ at $I$.
The normalized strength for a given transition
coincides with the normalized
transition probability or the branching ratio
for the decay from
$\alpha$ at $I$ to $\alpha'$ at $I-2$,
neglecting the $E_\gamma^5$ factor.

In the experiments, only
strong transitions are observed as discrete peaks in the
gamma-ray spectra  and the
rest of the transitions shows up as quasi-continuum spectra
which contain transitions summed over many states.
For such situation, it is useful to represent the E2
transition properties by means of the distribution function of the
strength, or the strength function.
The strength function  for the stretched E2 decay
from the levels at $I$ to the levels at $I-2$ is given
by
\beq
S_1(E_\gamma) = \sum_{\alpha\alpha'}
S_{\alpha I,\alpha' I-2} f_{\alpha I} \delta (E_\gamma - E_{\alpha I} +
E_{\alpha' I-2}),
\label{S1}
\eeq
where $f_{\alpha I}$ is the feeding probability of the level
$\alpha$ at $I$.
It becomes sometimes useful to define a strength function
\beq
S_{1,\alpha}(E_\gamma) = \sum_{\alpha'}
S_{\alpha I,\alpha' I-2} \delta (E_\gamma - E_{\alpha I} +
E_{\alpha' I-2}),
\label{trans}
\eeq
for stretched E2 gamma-decays from a given specific level $\alpha$ at $I$.
Average of $S_{1,\alpha} $ weighted with the feeding
probability is equivalent to the
strength function defined by Eq.(\ref{S1}).

\section{Results for \Yb}\label{sec:results}

\subsection{Mixing of $n$p-$n$h configurations}\label{sec:diag}

The calculations presented below were carried out
for the deformed rare-earth nucleus \Yb.
For this nucleus
there exist experimental data from the analysis of  quasi-continuum
gamma-spectra as well as  data from discrete-peak spectroscopy
identifying the rotational bands up to around $I \sim 40$
\cite{Fitz,Oliveira}.
The potential energy surface in this nucleus
has a  stable  minimum at prolate deformation
$\eps \sim 0.25$ up to $I \sim 60$ \cite{Werner-Dudek,
Bengtsson-Ragnarsson, Andersson}, and
many of the observed rotational bands are described as
independent particle excitations in the cranked mean-field
\cite{Oliveira,Bengtsson-Ragnarsson}.
This validates
the basis assumption of the present model.
We have also made calculations for other nuclei near \Yb obtaining
similar results.

\begin{figure}
\centerline{\psfig{figure=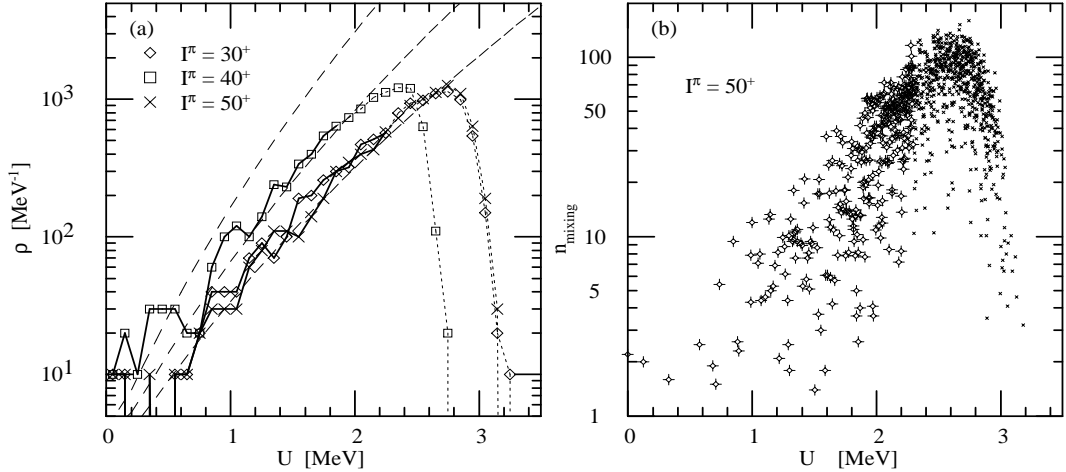,width=14cm,angle=-90}}
\caption{\label{fig3}
The level density
 for the calculated energy eigenstates with $I^\pi=30^+,40^+,50^+$
in \Yb is plotted in the left panel  as a function of
the excitation energy measured from the lowest state with same $I^\pi$.
The Fermi gas level density (long dashed lines) is also plotted
with the level density parameter $a=A/8,A/10,A/12$ MeV${}^{-1}$.
The thick lines cover the energy region containing
the lowest 300 levels while the short dashed lines continue above.
In the right panel, the mixing number $n_{mixing}$ is plotted
for the calculated $I^\pi=50^+$ states. The lowest 300 states
are labeled with a thick asterisk.
}
\end{figure}

The shell model diagonalization is done separately at
each $I^\pi$ for $I=20-61$. The energies of the  levels calculated
within  the lowest 10$^3$ basis states  cover an  interval of
$\sim 3$ MeV. The level density of these states
is plotted in Fig.\ref{fig3} for a few  spins as a function
of the excitation energy $U$ measured from the lowest
state for a given $I^\pi$. It is compared
with the Fermi gas level density with fixed signature and parity,
appropriate for the cranked mean field
\cite{Level-density},
\beq
\label{fermtot}
\rho_{FG}(U)=\frac{\sqrt{\pi}}{48}a^{-\frac{1}{4}}
U^{-\frac{5}{4}}\exp{2\sqrt{aU}}
\eeq
where  $a$ is the level density parameter.
It is seen that the calculated level density increases
exponentially as a function of the excitation energy, and
follows approximately
the Fermi gas formula
up to about 2.5 MeV above the lowest state. The level density
parameter fitting the calculated level density is about
$a \sim A/10$ MeV${}^{-1}$. This agrees with the standard estimate
\cite{Bohr-Mottelson} based on the
average single-particle level density in the
harmonic oscillator model.
The decrease of the level density
at $U \gesim 2.5$ MeV arises from the truncation of
the basis states.

As a consequence of the configuration mixing caused by the
residual interaction, the energy eigenstates $\{|\alpha\rangle\}$ are
admixture of many unperturbed states. In other words,
the unperturbed states are spread over the energy eigenstates.
The mixing is strongly
dependent on the internal excitation energy $U$
of the states since the level density
increases exponentially with $U$.
A measure of the configuration mixing can be defined
by means of a quantity
\beq\label{mixnum}
n_{\rm mixing}(\alpha I) = \left(\sum_\mu  \left| X_\mu^{\alpha}(I)
\right|^4 \right)^{-1},
\eeq
which counts effectively the number of basis states $\mu's$
participating to form an energy eigenstate
$\ket{\alpha}=\sum_\mu X^\alpha_\mu \ket{\mu}$.
Note that, when all the states $\mu$ have the same 
probability $\left|X_\mu^\alpha\right|^2$, $n_{\rm mixing}$
gives the exact number of admixed states. 
This quantity, called the mixing number, is plotted in Fig.\ref{fig3}(b)
for a typical case at $I^\pi = 50^+$ as a function
of the excitation energy $U$ measured from the yrast state.
For the states near
the yrast ($U \lesim 1$ MeV), the mixing number is less than about
3, which  indicates that each of these states is composed of 
a few dominant configurations. As $U$
increases, the mixing number increases steeply.  For the states at
$U \sim 2$ MeV, several tens of $n$p-$n$h
configurations contribute to form an
energy eigenstate  having  rather complex wave functions.
It is also to be noticed that the mixing number fluctuates
significantly state by state. There are several states with
low mixing number $n_{mixing} < 3$ even at $U = 1.5-2$ MeV
where most of the others  are strongly mixed
with $n_{mixing} \sim 10$.

The effect of the truncation is visible both in Fig.\ref{fig3}(a) and (b)
for the levels at $U \gesim 2.5$ MeV. We find that the lowest 300 levels
located at $U \lesim 2.3$ MeV
are stable against the truncation as far as
the level density and the gross features of the mixing number
are concerned.

\subsection{Onset of rotational damping}\label{sec:levels}

\begin{figure}[t]
\centerline{\psfig{figure=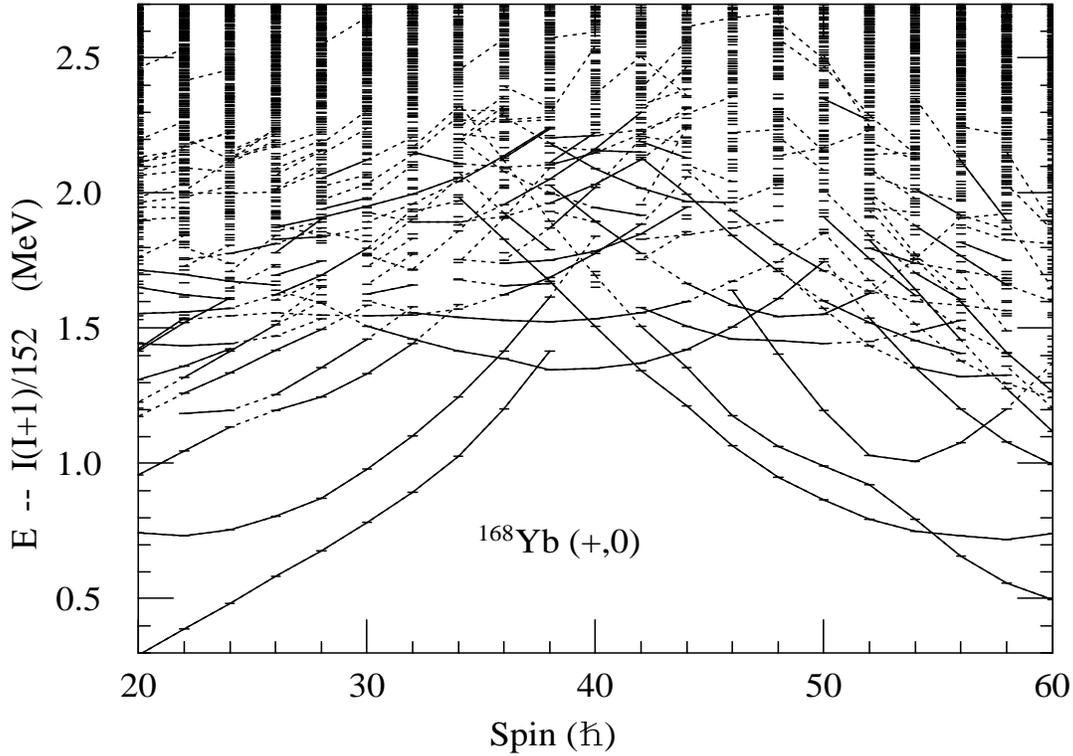,width=14cm,angle=-90}}
\caption{\label{fig4}
The calculated levels in \Yb with
$(+,0)$ are shown with small horizontal bars.
A reference energy
$I(I+1)/2J$ with $J=76$ MeV${}^{-1}$
is subtracted. The stretched E2 transitions which have
the normalized strength $S_{\alpha I, \alpha' I-2}$ larger than
0.707 are plotted with solid lines connecting  initial and final
levels of the transitions. Transitions with normalized
strength between 0.5 and 0.707 are plotted with dashed lines.
}
\end{figure}

The calculated energy levels
are plotted in Fig.\ref{fig4} with little
horizontal bars
for  $(+,0)$ states
(positive parity and even spin)
in \Yb. In this picture, a reference rotational energy is
subtracted from the calculated energy so that the vertical axis
represents roughly the internal excitation energy of the nucleus.
The solid lines connecting the energy levels
represents strong E2 transitions.
We use the
convention that the E2 decay is ``strong'' if the
associated normalized strength satisfies
$S  > 1/\sqrt{2} = 0.707$.
Weaker transitions with  $0.5 < S < 0.707$
are displayed  with dashed lines.
It is seen that the strong E2 transitions (solid lines)
form  sequences of levels which
are aligned regularly along a parabola like curve. Such
sequences of levels represent rotational band structures.
Most of the rotational bands lie
in the region near the yrast, while the levels at higher excitations
energy as a rule do not form band structures.

\begin{figure}[t]
\centerline{\psfig{figure=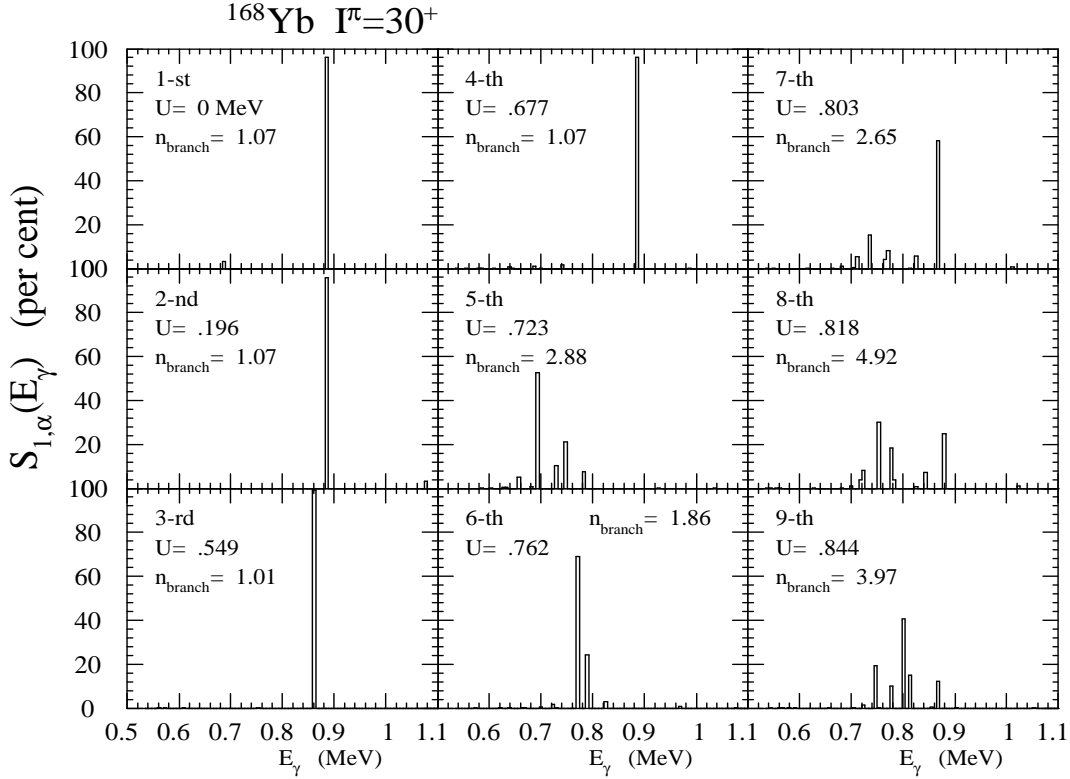,width=14cm,angle=-90}}
\caption{\label{fig5}
The distribution $S_{1,\alpha}(E_\gamma)$
of the stretched collective E2 decays
from the lowest 9  levels with
$I^\pi = 30^+$. The branching number $n_{branch}$ defined
by Eq.(\ref{eq:branch}) for the E2 decay and the relative excitation
energy $U$ (in unit of MeV) are  put for each level.
The bins for the transition gamma energy $E_\gamma$ have
a width of 6 keV.
}
\end{figure}

\begin{figure}
\centerline{\psfig{figure=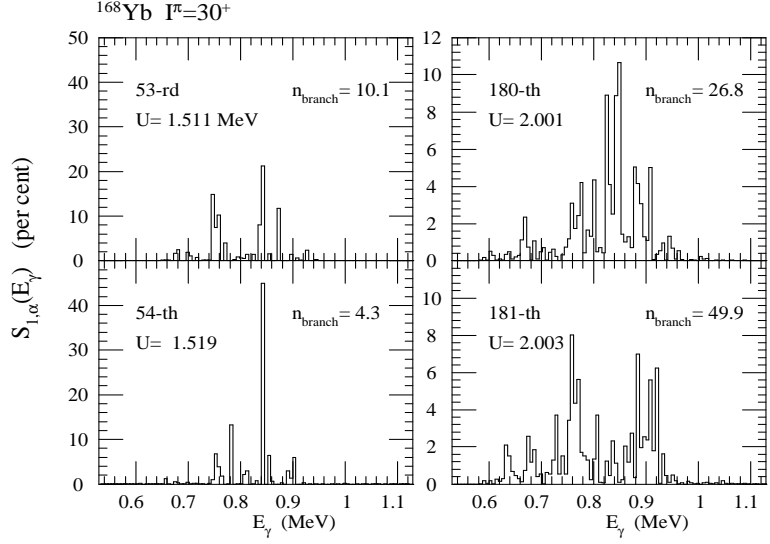,width=10cm,angle=-90}}
\caption{\label{fig6}
The strength distribution $S_{1,\alpha}(E_\gamma)$
for the stretched collective E2 decays
from the 53-rd, 54-th, 180-th, and 181-st excited levels
with $I^\pi =30^+$.
}
\end{figure}

To indicate the E2 transition properties more precisely, we
calculate distribution of the E2 transition strengths
$ S_{1,\alpha}(E_\gamma)$, Eq.(\ref{trans}), defined
for gamma-decays from individual levels $\alpha$ at $I^\pi$.
They are  shown in Fig.\ref{fig5}
for the lowest levels  with $I^\pi =30^+$.
The E2 strength associated
with the first  $30^+$ level (which is the second lowest at spin $30$)
is concentrated in a single
component feeding the $28_1^+$ level with strength exhausting more
than $95\%$ of the total strength. The $30_2^+$, $30_3^+$ and $30_4^+$
levels show essentially the same E2 distribution except slight
difference in the gamma-ray energy for the dominant transition.
The level $30_5^+$, with an internal excitation
energy of  about 700keV, displays a
completely different E2 strength
distribution, being fragmented over several transitions, each
of which carries  a rather weak strength.
The
E2 strength associated with the decay from
the $30_6^+$, $30_7^+$, $30_8^+$
and $30_9^+$ levels shows a similar fragmentation.
The fragmentation of the E2 strength increases as the excitation energy
increases. Figure \ref{fig6} displays the quantity
$S_{1,\alpha}(E_\gamma)$
associated with
the levels
$30_{53}^+$ and $30_{54}^+$ lying at $U \sim 1.5$ MeV and for
the levels $30_{180}^+$ and $30_{181}^+$ lying at $U \sim 2.0$ MeV.
At $U \sim 1.5$ MeV, the E2 strength distribution has about ten
branches, while the number of branches becomes much larger at
$U \sim 2.0$ MeV. The E2 strength is distributed within the range
$E_\gamma \sim 0.7 - 1.0$ MeV, with centroid  at
$E_\gamma \sim 0.85$ MeV and with a width of about 150 keV.

\begin{figure}
\centerline{\psfig{figure=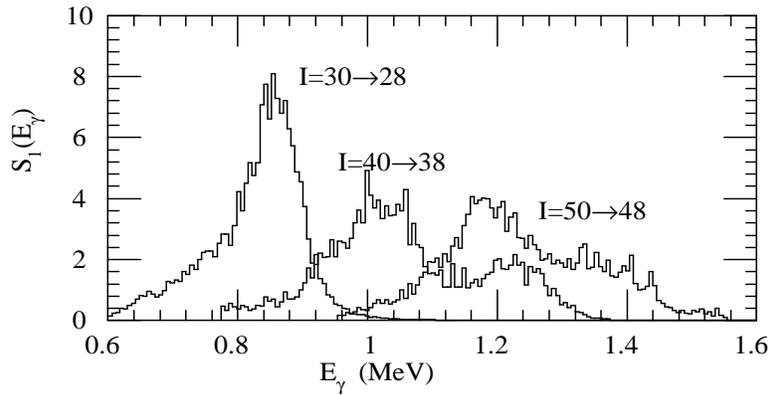,width=10cm,angle=-90}}
\caption{\label{fig7}
The  strength function $S_1(E_\gamma)$
 for the E2 strength for the
decays $I=30\rightarrow28$, $I=40\rightarrow38$ and $I=50\rightarrow48$,
summed  over the lowest 200 states for both parities with a
constant feeding probability $f_{\alpha I}$.
}
\end{figure}

The fragmentation of the E2 strength function
is the rotational damping phenomenon
\cite{Lauritzen}.
Averaging
the E2 strength distributions over many states (Fig.\ref{fig7}) produces
smooth profile for the strength distribution.
The centroid of the
distribution increases with the spin as expected from
the relation  $\langle E_\gamma \rangle \sim 2
\hbar\omega_{rot} \sim 2 I /J$,  $J$ being the average
moment of inertia. The width of the distribution is
about 100 keV for $I \sim 30$,  and 250 keV for $I \sim 50$.
This approximately agrees with the estimate
for the rotational damping width
$\Gamma_{rot} = 4\Delta\omega =0.14 (I/40)U^{1/4}$  MeV
in Ref.\cite{Lauritzen}.
To be more precise, however, the strength distributions
are not necessarily represented by a simple 
Lorentzian or Gaussian, but exhibit structures which
reflect specific alignment properties of the underlying
single-particle orbitals.
In this particular nucleus, there exist two significantly aligned
proton orbitals near the Fermi surface, $\pi h_{9/2}$ and
$\pi i_{13/2}$ (see Fig.\ref{fig1}).
These aligned proton orbitals cross the $Z=70$ Fermi surface
at $\omega \sim 0.4$ MeV or $I \sim 40$.
The peak
at $E_\gamma \sim 1.0$ MeV of the
$I=40\rightarrow38$ strength distribution
consists mainly of the transitions
involving  one of the aligned $\pi h_{9/2}$ and
$\pi i_{13/2}$ orbitals  while the little bump at
$E_\gamma \sim 1.2$ MeV arises from the
components containing neither $\pi h_{9/2}$ nor
$\pi i_{13/2}$.  An effect of the aligned proton orbits
is also seen in Fig.\ref{fig4} as changes
in slopes of the rotational band structures near the yrast at $I\sim 40$
(see also the later discussion).

\begin{figure}
\centerline{\psfig{figure=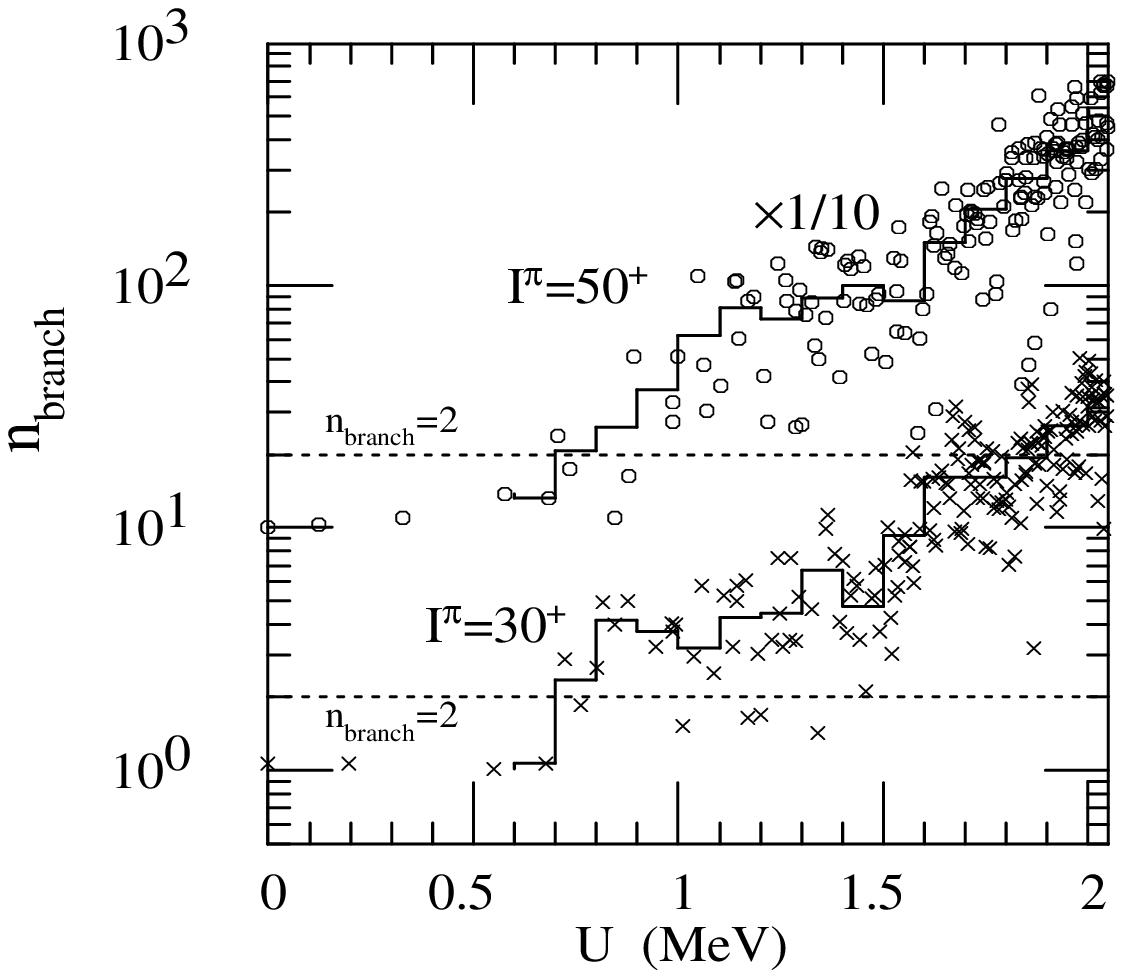,width=10cm,angle=-90}}
\caption{\label{fig8}
The E2 branching number $n_{branch}$
for the $I^\pi =30^+$ (crosses), $50^+$ states (circles)
as a function of the excitation energy $U$ of the states measured
from the lowest one. The histogram shows $n_{branch}$  averaged for
energy bins.
}
\end{figure}

It may be useful to define the onset energy
where the rotational damping sets in.
In order to quantify the onset of rotational damping,
let us utilize the  branching number \cite{MatsuoPT,Matsuo93,MatsuoNPA}
\beq \label{eq:branch}
n_{\rm branch}(\alpha)\equiv\left(\sum_\beta S_{\alpha I,
 \beta I-2} ^2\right)^{-1}
\eeq
which counts effectively the number of the E2 branches for
decays from a level $\alpha$ at $I$ to levels at $I-2$.
This quantity is analogous to the mixing number
introduced in Eq.(\ref{mixnum}).
Figures \ref{fig5} and \ref{fig6} also show the branching number
for each level. As seen in Fig.\ref{fig5}, the onset of damping
can be characterized by a condition $n_{branch}>2$ implying an
E2 decay with more than 2 branches. The dependence of the
branching number with excitation energy
is depicted in Fig.\ref{fig8}. The branching
number increases exponentially with internal excitation energy.
With the criterion $n_{branch}>2$ for the  onset of  rotational
damping, the onset energy is read from Fig.\ref{fig8} to be
 $U \sim 800$ keV above yrast.

It should be emphasized that, although the onset energy thus defined
tells approximately where the rotational damping sets in, the
transition from the region of rotational bands to the region of
rotational damping does not take place sharply at the onset energy, but
rather develops gradually as the excitation energy increases.
In fact, Fig.\ref{fig4} shows
presence of many {\it short rotational bands}
for which the strong E2 transitions (solid or dashed lines) continue
for only a few to several
steps. They lie from  the border  region $U \sim 1$ MeV of
the onset of rotational damping  up to a region of much higher
level density  with
$U \sim 1.5$ MeV. Here the short band structures  are
surrounded by  levels which do not have any strong transitions.
This indicates that the
rotational band structures partly remain even in the region of the
rotational damping. Presence of such scars of rotational bands
\cite{Scar} is also displayed in Fig.\ref{fig8}, from which
it is seen that
there exist levels whose branching
number $n_{branch}$ is smaller
than 2 or 3 even at high excitation energy $U \sim 1.5$ MeV.
It is noticeable that the E2 strength distribution in the transition region
is fluctuating quite irregularly from state to state as
the  Figs.\ref{fig6} and \ref{fig8} indicates.
Two panels in the left (right) hand side show quite different
fine structures, even if  they are produced from neighboring levels
$30_{53}^+$ and $30_{54}^+$ ($30_{180}^+$ and $30_{181}^+$),
corresponding to excitation energies which are the same within
just 8 (2) keV.


\subsection{Number of rotational bands}\label{sec:nband}

Since the rotational band structures do not survive
in the region of high internal excitation energy,
there exists only a finite number of
discrete rotational bands in a single nucleus. This feature can be
utilized to study experimentally the onset of rotational
damping. Through
the fluctuation analysis method \cite{FAM-168Yb,FAM-PR}
it is possible to extract an effective number of gamma decay paths
from the double-coincident
$E_\gamma \times E_\gamma$ spectrum. When this method is
applied to the ridge structures in the spectrum, the
effective number of paths essentially corresponds to the
number of discrete rotational bands.

\begin{figure}
\centerline{\psfig{figure=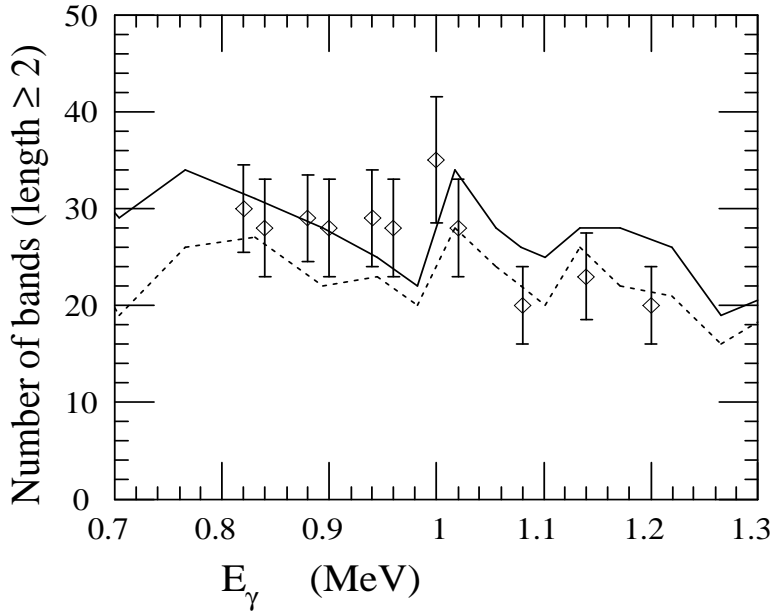,width=10cm,angle=-90}}
\caption{\label{fig9}
The calculated number of bands with length more  than or equal to
2 is compared with the experimental effective number of paths
\cite{FAM-PR} extracted
from the first ridge of $(E_{\gamma 1},E_{\gamma 2})$ spectra in
\Yb. The solid line is calculated with the criterion
$n_{branch}<2$ while the dashed line is defined
with
$S >0.707$.
The horizontal axis denotes the average gamma-ray energy
$E_\gamma = (E_{\gamma 1}+E_{\gamma 2})/2$.
}
\end{figure}

Let us  define a corresponding quantity
within the framework of the present calculation
to be able to compare our results with the experimental findings.
A straightforward definition is just to count the
number of strong E2 transitions at a given spin with strength $S$
larger than a given threshold $S_{thr}$, for which one
may use $S_{thr}=1/\sqrt{2}=0.707$ to define levels
forming discrete rotational bands as adopted in plotting Fig.\ref{fig4}.
Keeping in mind the
previous analysis for the onset of damping,
another definition may also be used in terms of the
branching number with criterion $n_{branch} <2 $
for levels forming rotational bands.
Note that these two criteria are not very different, since each state
which has a very strong transition with $S>0.707$ necessarily has
a branching number $n_{branch} <2$.
The fluctuation analysis method is often applied to
the first ridge located within the interval
$E_{\gamma 1} - E_{\gamma 2} = \pm 4/J$
in the  $E_\gamma \times E_\gamma$ spectra ($J$ being the moment of
inertia), which
is formed by two consecutive E2 gamma-rays.
Correspondingly we count the number of
rotational bands which satisfy the above criteria 
at least over two steps $I+2 \rightarrow I \rightarrow I-2$ of E2 decays.
In practice, we count all the states  which satisfy
$n_{branch} < 2$ or $S >0.707$
for both  decaying transitions  and feeding ones, and
sum the numbers from four sets of parity and signature
$I^\pi=I_0^+,I_0^-,(I_0+1)^+,(I_0+1)^-$ for a given representative
spin $I_0$. The calculated number of bands can be plotted
as a function of spin $I_0$ or the average gamma ray energy
$E_\gamma =(E_{\gamma 1} + E_{\gamma 2})/2$.
Figure \ref{fig9} shows the calculated number of bands
using the two criteria and compares them with the experimental
effective number of paths associated with the first ridge. The two criteria
give essentially the same number of bands around 30. It is noticed
that the theoretical calculations and the data agree quite well
in all the $\gamma$-ray energy range $E_\gamma \sim 0.8 - 1.2$ MeV
corresponding to spins $I \sim 28 - 48$.

More accurate comparison between the theory and the experiments
requires evaluation of the
feeding probabilities.
This is possible by
making a simulation of the whole gamma decay
cascades combining microscopically calculated
levels and E2 transitions with statistical description of
E1 decays since
the present model can describe levels 
up to about 2 MeV above yrast line at  high spins,
where most of the gamma decay cascades of a warm rotating
compound nucleus proceed. 
Such a microscopic
simulation has been developed recently\cite{Bracco}.
{}From the  $E_\gamma \times E_\gamma$ spectrum produced
from the  simulation, the effective number of
paths is extracted in the same way as the experimental
analysis. In this manner, we can make comparison
which does not depend on the particular choice of the
threshold for $S_{thr}$ or $n_{branch}$ used
for the definition of the number of bands.
It is found that the effective numbers of paths both in
the simulation and in the experiments for \Yb agree quite
well \cite{Bracco}.

\begin{figure}[t]
\centerline{\psfig{figure=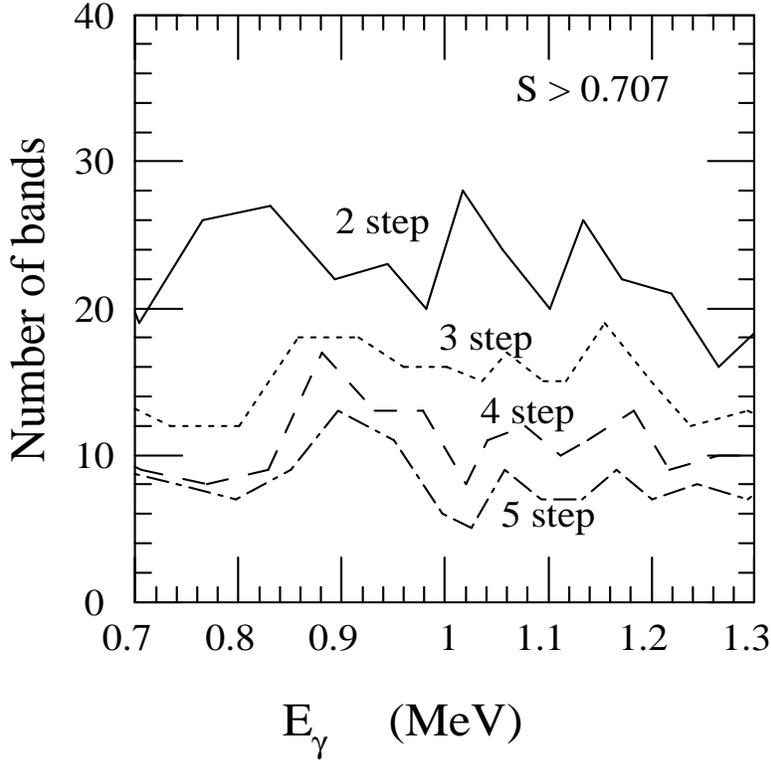,width=10cm,angle=-90}}
\caption{\label{fig10}
The calculated number of bands with length more  than or equal to
2, 3, 4 and 5 steps, plotted with solid, dotted, and dot-dashed lines,
respectively.
In this figure, the criterion
$S >0.707$
is used to define the rotational bands.
The horizontal axis denotes the average gamma-ray energy
$E_\gamma = (E_{\gamma 1}+E_{\gamma 2})/2$.
}
\end{figure}

As discussed above, the calculations predict the presence of
scars of rotational
bands, that is, the ``short'' rotational band structures
connected by strong E2 transitions for only
a few decay steps.
It may be possible to check this feature in the
experiments by looking into the second and higher order
ridges in the $E_\gamma \times E_\gamma$ spectra.
As the two consecutive gamma-rays from a
rotational band fall on the first ridge at
$E_{\gamma 1} - E_{\gamma 2} = 4/J$, two gamma-rays of the
first and last decay steps in $n$ consecutive
transitions should fall on the $(n-1)$-th ridge at
$E_{\gamma 1} - E_{\gamma 2} = 4(n-1)/J$.
Thus, for example,  the effective number of path extracted from the
second ridge tells the number of rotational band structures which
continue for at least three consecutive  steps. Since
the rotational bands continuing only for two steps do not
contribute to the second ridge, we may expect the effective number
of paths for the second ridge to be smaller than that for the first 
ridge.
This is illustrated  in Figure \ref{fig10}.
In this figure, we count the number of rotational bands
defined by the criterion $S > 0.707$ for various lengths of
decay steps. In plotting the number of rotational
bands, we used the average gamma-ray energy for the two gamma rays,
$I+2 \rightarrow I$ and $I-2n+2 \rightarrow I-2n$, for the
rotational bands with $\ge n$ decay steps. The calculation indicates
that a considerable part of the first ridge comes from
the scars of rotational bands at "high" heat energy. It also suggests
that the effective number of path in the higher order ridges may
be significantly smaller than in the first ridge.
In fact, the experimental data for the higher order ridges also indicate
this tendency \cite{Erice}. In order to make a 
quantitative comparison, however, a more careful analysis using e.g.,  the
microscopic simulation of Ref.\cite{Bracco} is required.

\subsection{$E_\gamma - E_\gamma$ correlation}\label{sec:egam-egam}

Since the quasi-continuum $E_\gamma \times E_\gamma$ spectrum displays the
characteristic ridge-valley structure, the shape of the
spectrum provides us with important information about the
rotational damping \cite{RPM,Leoni}.
The quasi-continuum analysis often deals with
projection of the $E_\gamma \times E_\gamma$ spectra upon
the $E_{\gamma 1} - E_{\gamma 2}$ axis.
A quantity which is related to the projected spectra is the
strength distribution function of the coincident two E2 transitions.
In particular, the two-gamma strength function for consecutive
E2 decays $I+2 \rightarrow I \rightarrow I-2$ is important
since it characterize the shape of the first ridge
as well as of the central valley. The two-gamma strength function
is given by
\beq
S_{2}(E_{\gamma 1} - E_{\gamma 2})=\sum_{\alpha\alpha'\alpha''}
S_{\alpha I+2,\alpha' I}S_{\alpha' I,\alpha'' I-2}
f_{\alpha' I}\delta(E_{\gamma 1} - E_{\gamma 2}
{}-E_{\gamma,\alpha \rightarrow \alpha'}
+E_{\gamma,\alpha' \rightarrow \alpha''})
\eeq
where
\beqa
E_{\gamma,\alpha \rightarrow \alpha'} & = E_{\alpha I+2} - E_{\alpha' I} \\
E_{\gamma,\alpha' \rightarrow \alpha''} & = E_{\alpha' I} - E_{\alpha'' I-2}
\eeqa
and the $f_{\alpha I}$ is the feeding probability of the level
$\alpha$ at spin $I$.

\begin{figure}[t]
\centerline{\psfig{figure=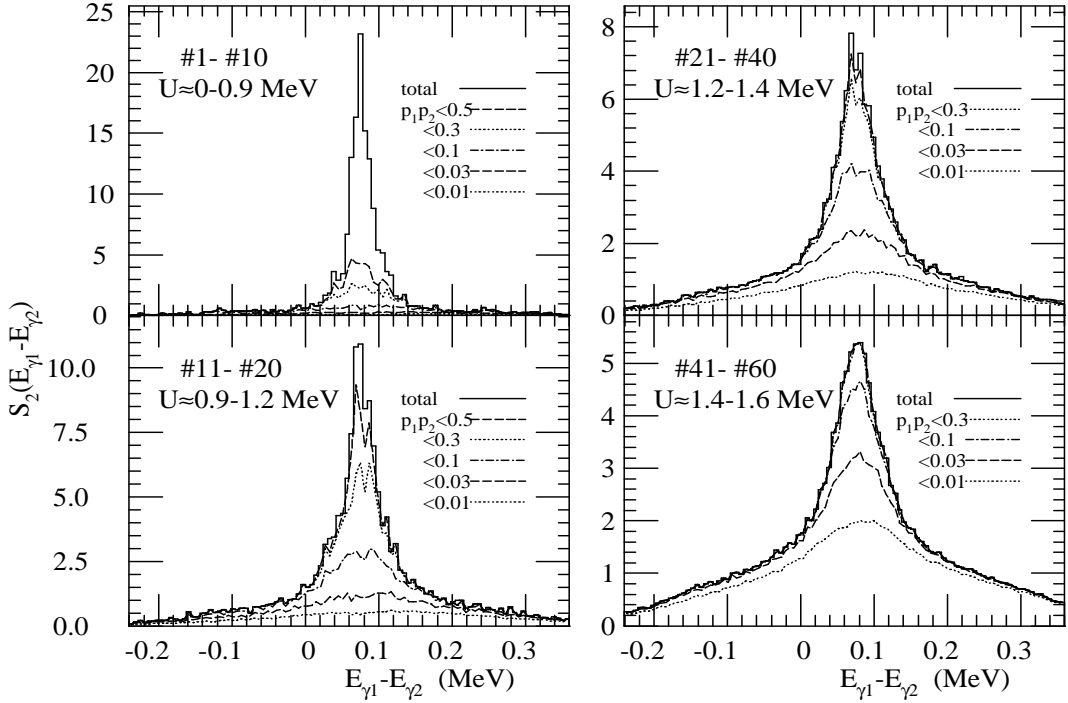,width=14cm,angle=-90}}
\caption{\label{fig11}
The two-step strength distribution $S_2(E_{\gamma 1}-E_{\gamma 2})$
of consecutive E2 transitions for different bins of
excited levels, defined for the lowest 10 levels, the 11-st to 20-th
levels, the 21-st to 40-th, and 41-st to 60-th
for each $I^\pi$. The strength is averaged  for the spins
$I=30-51$ in order to get enough statistics to produce smooth profile.
The strength distribution is also subdivided with respect to
the product strength $p_1p_2=
S_{\alpha I+2,\alpha' I}S_{\alpha' I,\alpha'' I-2}$ of
individual transitions.
The approximate excitation energy of the bin measured from yrast
is put for each bin for the sake of reference.
}
\end{figure}

The calculated two-gamma strength function
$S_{2}(E_{\gamma 1}-E_{\gamma 2})$ is depicted in Figure\ \ref{fig11}.
Here an average is taken over the levels within a
spin interval $I=30-51$.
In order to study the excitation
energy dependence, the sum over $\alpha'$
is divided into bins including 10 or 20 levels
for each $I^\pi$.
The feeding probability
$f_{\alpha' I}$ is put equal for all the levels.
The lowest bin $(\#1-\#10)$ covers approximately the energy region
$U = 0-0.9$ MeV, thus it mostly contains the E2 transitions associated
with the rotational band structures near the yrast line. The higher
bins cover the region above
the onset of rotational damping.

For the lowest bin, the most characteristic feature is the
presence  of  a sharp peak located at
$E_{\gamma 1} - E_{\gamma 2} \sim 70$ keV
with a width of about 30 keV, which corresponds to the
first ridge observed in the $E_{\gamma} \times E_{\gamma}$ spectrum.
In order to show the contents of the peak, the two-gamma
strength function $S_2(E_{\gamma 1} - E_{\gamma 2})$ is subdivided by putting
cuts on the product strength  $S_{\alpha I+2,\alpha' I}
S_{\alpha' I,\alpha'' I-2} \equiv p_1 p_2 $ of each transition
$(\alpha,I+2) \rightarrow (\beta,I+2) \rightarrow (\gamma,I-2)$.
The sharp ridge in the lowest bin consists mostly of the
strong transitions satisfying $p_1 p_2 >0.5$. This indicates
that the sharp ridge is formed by
the strong transitions ($p>1/\sqrt{2}$) associated with
the rotational band structures.
The peak position of the sharp ridge is related to
an average  dynamic moment of inertia $J$ of these bands through
$E_{\gamma 1} - E_{\gamma 2}=4/J$, and the peak width originates
from fluctuations in the moment of inertia among different bands.

For the higher bins ($\#11-\#20,\#21-\#40,\#41-\#60$),
the profile of the two-gamma strength function is quite
different from that of the lowest bin in many respects.
Note that there is only a little contribution of
strong E2 transitions ($p_1p_2 >0.5$) associated with
the rotational band structure.
This is because the rotational damping causes
fragmentation of the E2 strengths with a spread
in gamma ray energy of order of $\Gamma_{rot}$.
Moreover, the spectrum shows a two component profile
with wide and narrow widths of about 300 and 80 keV.
The intensity of the wide component 
increases with the average energy of the
bins.
This component is easily understood as a consequence of
the rotational damping.
If two consecutive transitions were
uncorrelated,
the two-gamma strength function $S_2(E_{\gamma 1} - E_{\gamma 2})$
would simply become a convolution
of two single-gamma strength functions $S_1(E_\gamma)$ for the
consecutive steps, producing a smooth distribution
with a width a
factor of 2 larger (assuming a Lorentzian shape)
than that of the single-gamma strength function
$S_1(E_\gamma)$,
which
displays a width of about 100-200 keV (see Fig.\ref{fig7}).

The narrow component has a
width of about 80 keV, whose  intensity decreases with the energy of the
bin.
The origin of the narrow component is visible in Fig.\ref{fig11} by
subdividing the strength function with respect to
the product strength $p_1 p_2$ of individual transitions.
It is  noticed that
the transitions building up the narrow component have
{\it relatively} large
strength compared to those composing the wide components.
Taking the third bin ($\#21-\#40$) as an example,
the narrow component consists of
transitions with $0.03 < p_1 p_2 < 0.3$
while weaker transitions $ p_1 p_2 < 0.03$ contribute only to
the wide component. These transitions in the
narrow component are much stronger than the average
strength $(p_1 p_2 \sim (\Gamma_{rot} \rho)^{-2} \sim 0.01)$ 
expected for damped transitions.

This analysis indicates that
the correlation
$E_{\gamma 1} - E_{\gamma 2}  \sim 4/J$
associated with the rotational band structure still remains
in the consecutive E2 transitions even after the rotational damping sets in and
that the narrow component has the same origin
as the ``scars'' of the rotational bands discussed above.
This suggests that the width of the narrow component
is related to  the spreading width which represents the extent
of the configuration mixing of the  rotational band structures.

\begin{figure}[t]
\hskip 1cm
\begin{minipage}[t]{6cm}
\psfig{figure=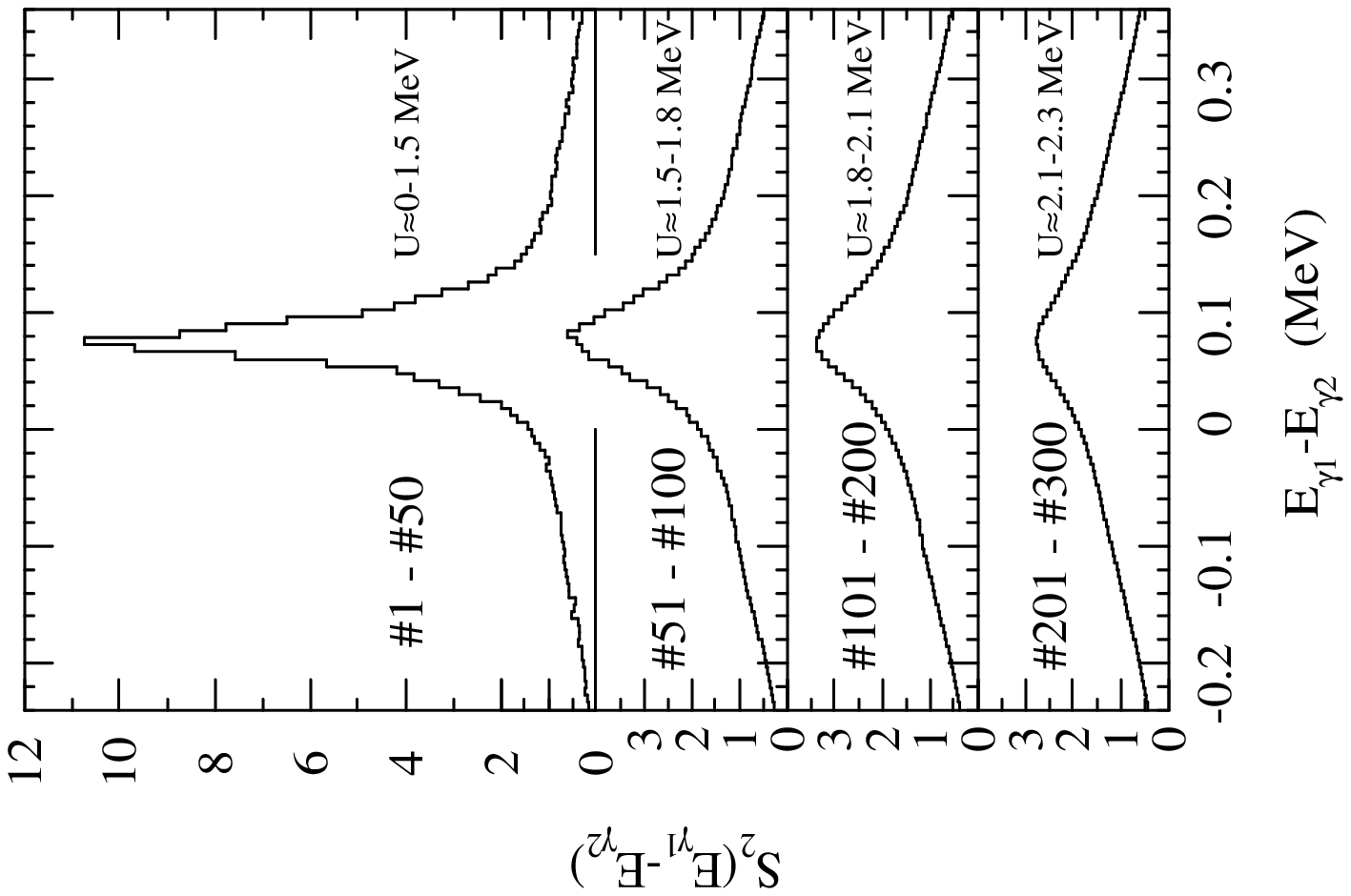,width=5cm,angle=-90}
\caption{\label{fig12}
Excitation energy dependence of
the two-gamma strength distribution $S_2(E_{\gamma 1}-E_{\gamma 2})$
for the energy bins covering the first to 50-th,
51-st to 100-th, 101-st to 200-th, and 201-st to 300-th lowest
levels at each $I^\pi$. The strength is averaged over spin
interval $I=30-51$.
}
\end{minipage}
\hskip 1cm
\begin{minipage}[t]{6cm}
\psfig{figure=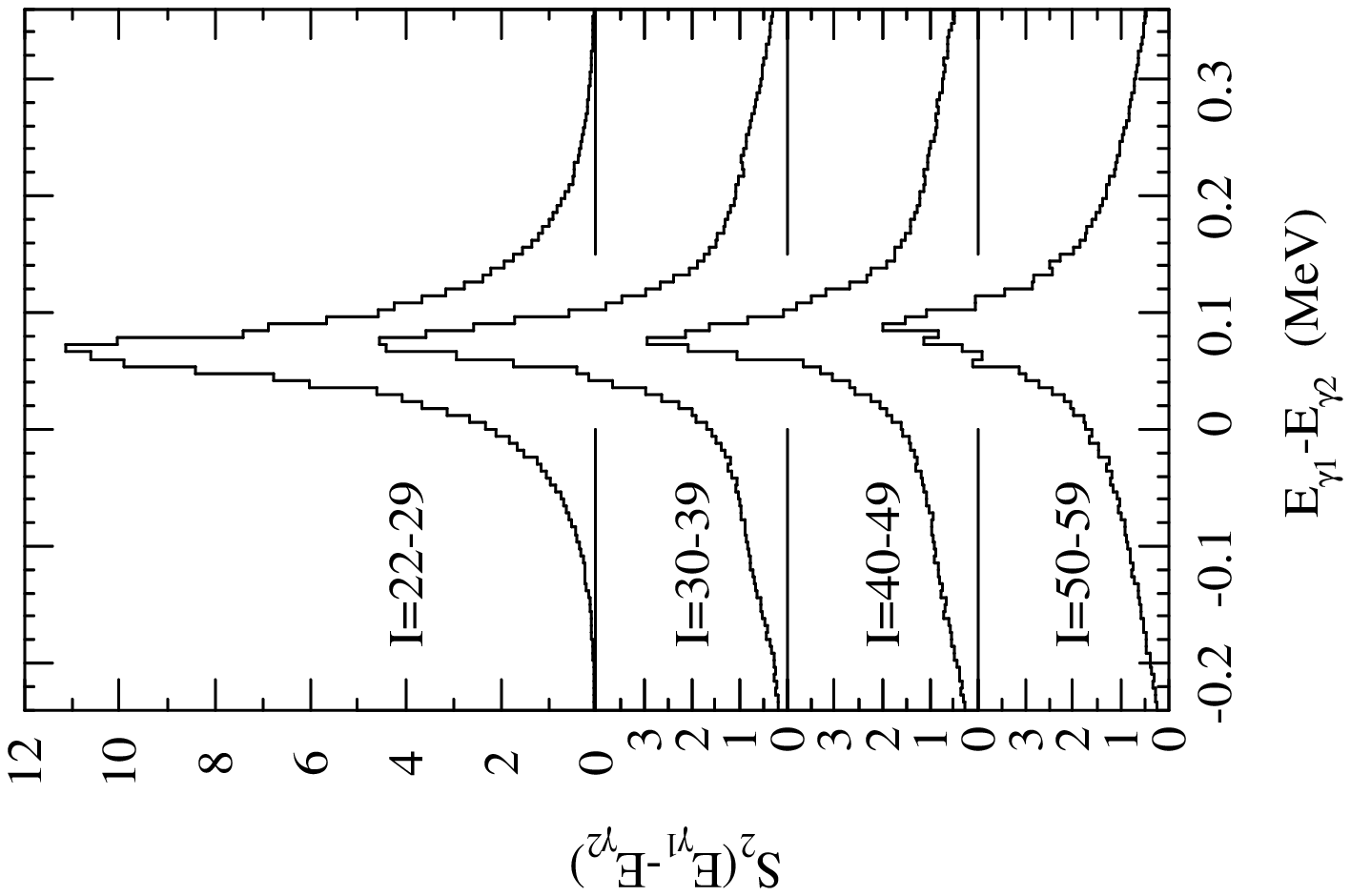,width=5cm,angle=-90}
\caption{\label{fig13}
Spin dependence of
the two-gamma strength distribution $S_2(E_{\gamma 1}-E_{\gamma 2})$
for the energy bin covering the lowest 100 levels at each
$I^\pi$. The strength distribution is shown for different
spin intervals; $I=22-29,30-39,40-49,50-59$.
}
\end{minipage}
\end{figure}

It is also noticed that
the narrow component is intense even at the fourth
bin ($\#41-\#60$) corresponding to excitation energy
$U \sim 1.5$ MeV above yrast.
It becomes less
intense as the excitation energy increases, and almost disappears 
at $U \gesim 2$ MeV as shown in Fig.\ref{fig12}.
The development of rotational
damping is quite gradual as
a function of $U$, and the transition region extends
from $U \sim 0.8$ MeV to $U \sim 2$ MeV above yrast line.
Interestingly, it is found from the statistical
analysis of energy level spacing and  individual E2 strength
\cite{Matsuo93,MatsuoNPA,MatsuoPT}
that the E2 strength shows the Porter-Thomas distribution
at $U \gesim 2$ MeV while it deviates significantly from the
random limit for $U \lesim 2$ MeV. The deviation from the Porter-Thomas
distribution and the presence of the narrow component are
related with each other since both originate from the presence of
strong transitions in the region of the rotational damping.

The $E_\gamma - E_\gamma$ correlation only displays a
weak dependence on angular momentum, 
as shown in Figure \ref{fig13}. The intensity of the narrow component
becomes larger for lower spins although there exist significant
narrow components even at very high spin ($I \gesim 50$).
The smaller the spin is, the smaller becomes the rotational damping
width (width of the wide component). When the
damping width becomes smaller,
the branching number and strength of the fragmented E2 transitions
becomes
smaller and larger, respectively. Thus strong transitions which
cause the narrow components become more dominant at lower spins.
On the other hand, the width of the narrow component is
not very dependent on spin. This also suggests
that the narrow width may be related to
the spreading width, which is essentially spin independent
in the present model.

In the experiments, the actual  $E_\gamma \times E_\gamma$ spectrum
is not simply represented by the two-gamma strength function
discussed above since it is  formed not only by
the consecutive E2 transitions but also by non-consecutive
E2's as well as E1 transitions all weighted by the
feeding probability. It is found, however,
by means of the simulation analysis \cite{Bracco} that
the two-component profile causes significant effects on the
spectral shape of the $E_\gamma \times E_\gamma$ spectra \cite{Leoni}.

\subsection{Near-yrast rotational bands}
\label{sec:low-lying}

The main aim of the present model is to
describe the internally excited  states  and
the rotational damping phenomena
in the warm region of the spectrum at very high spin. 
However, it is interesting to study how
the two-body residual interaction affects the rotational
bands near the yrast line as 
compared to  the standard cranked mean-field calculations
which are often used to describe those states.

\begin{figure}[t]
\centerline{\psfig{figure=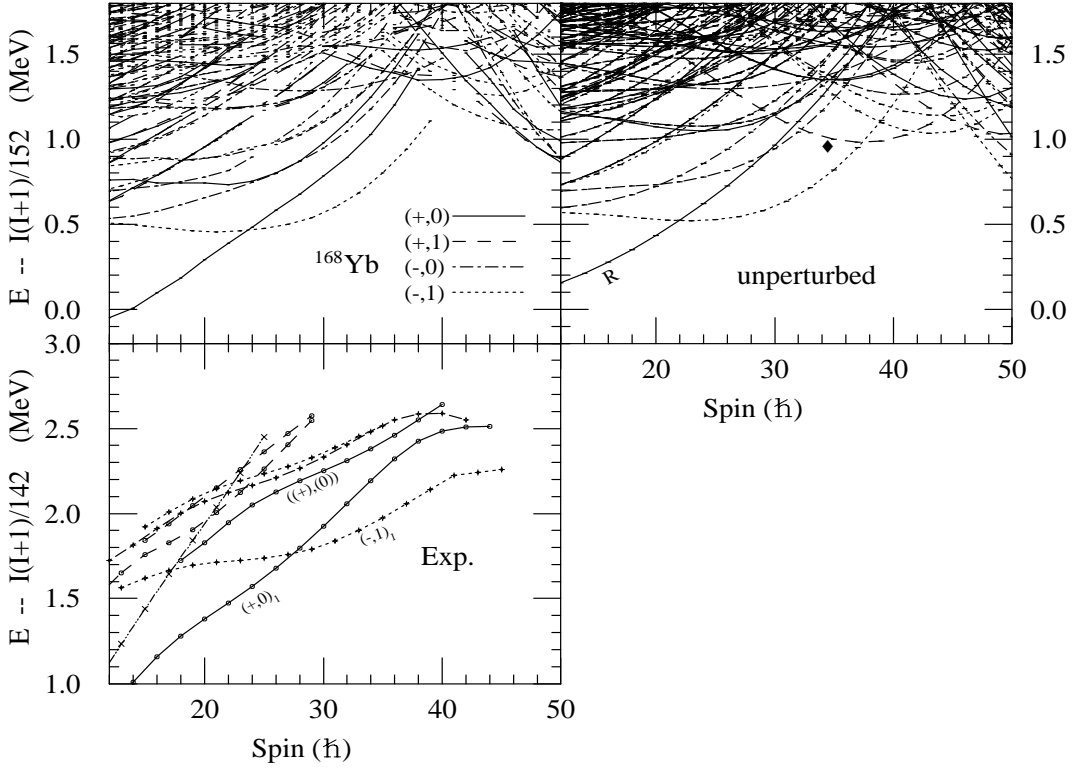,width=14cm,angle=-90}}
\caption{\label{fig14}
The rotational band structures near yrast line in \Yb.
The calculated band structure is shown in the left-top panel
in a similar way to Fig.\ref{fig4} subtracting the 
reference rotational energy $I(I+1)/152$ MeV. The right-top panel
shows the unperturbed rotational bands calculated without the
residual two-body force. See text for marks.
The left-bottom panel shows the observed rotational bands
subtracting the reference energy $I(I+1)/142$ MeV
\cite{Fitz,Oliveira}.
}
\end{figure}

The calculated and observed rotational bands
are compared in
Figure \ \ref{fig14}. The rotational bands
below spin $I \sim 20$ are not reproduced well
in the present model.
This is because the
present calculation  does not give a satisfactory description
of the pairing correlations,
which are important for the rotational bands at low spins.
Pairing correlations may also account for the slightly lower
moments of inertia of the experimental bands in the
angular momentum region around $I \sim 20$ to 40. This is
taken into account by using a reference moment of inertia in
Fig.\ref{fig14}, which is chosen smaller by about $10 \%$ for
the experimental bands than for the theoretical ones.
In the higher spin region with $I \gesim 30$, the
model reproduces fairly well the overall features of the
lowest few bands near the yrast line; above the crossing at $I \sim 25$,
the $(-,1)_1$ band becomes the yrast band, leaving the
$(+,0)_1$ as the second lowest, both in the theory and in the
experiment. 
The dominant configuration in the  $(-,1)_1$  band is neutron
1p1h excitation $(\nu[521]1/2)^1(\nu[642]5/2)^{-1}$ relative
to the reference configuration (the $(+,0)$ band marked with R
in Fig.\ref{fig14}). The third lowest $((+),(0))$ band in the experiments
seems to correspond to the $(+,0)_2$ bands in the model, whose
dominant configuration is the neutron excitation
$(\nu[521]1/2)^2(\nu[642]5/2)^{-2}$.
Furthermore, both the model and the experiment show the band
crossings at $I \sim 40$. In the calculation, the crossings
involve the aligned proton orbits
$\pi h_{9/2}$ and/or $\pi i_{13/2}$
(See Fig.\ref{fig1}). They are, however, slightly sharper
than those in the observed bands. This discrepancy
also may be attributed to the insufficient pairing correlation
in the present calculation
(Note that Ref.\cite{Fitz} discusses  the smoothness of
the crossing in connection with
the proton pairing).

In order to show the effects of the
interaction, we compare in Fig.14 with
the unperturbed rotational bands which are obtained without
the residual interaction.
It is noticed that the
$(+,0)$ configuration taken as the reference (marked with R)
gains  energy due to the correlation caused by the
residual interaction, which includes a part of the pairing effect.
On the other hand, the (+,1) unperturbed rotational band  marked with
a diamond is pushed up by the residual interaction. This band has
configuration $(\nu[521]1/2)^1 (\nu[642]5/2)^{-1}$ for neutrons
and $(\pi h_{9/2})^1(\pi[411]1/2)^{-1}$ for protons with respect to
the reference. Since the neutron and the proton configurations
have different spatial density distribution (oblate and
prolate along the symmetry axis, respectively), the residual
SDI force makes this configuration energetically unfavored.
The residual interaction gives, at least in this case, an
overall improvement of the description of the
near-yrast rotational bands compared to the unperturbed
rotational bands with pure independent particle configuration.

\section{Residual interactions}\label{sec:resint}

\subsection{Interaction strength}\label{sec:intstr}

Since the rotational damping
is controlled predominantly by the configuration mixing caused
by the residual interaction, it is interesting to
examine the dependence on
the two-body residual force.
It is also noticed that there
is some ambiguity in the strength of the
SDI (the literature value ranges as $V_0 \sim 20/A - 30/A$
MeV, see Appendix B).  To examine
the dependence of the results  on the SDI strength,
we performed calculations with various SDI strengths ($V_0 = 14/A,
20/A, 35/A, 50/A$ MeV).
We also  check how the results depend on the
specific features of the SDI that
emphasize the interaction at
the nuclear surface and give large matrix elements for
the single-particle orbits near the Fermi surface.
For that purpose, we also perform a calculation with
a {\it volume} delta force
\beq
v(1,2) = - v_\delta \delta(\vec{x}_1 - \vec{x}_2)
\eeq
which does not have the surface effects.
For the strength of the volume delta interaction, we
used $v_{\delta,nn}=v_{\delta,pp} = 340 $
fm${}^3$MeV and $v_{\delta,pn} = 500$ fm${}^3$MeV
taken from Ref.\cite{Delta}, which are evaluated
from comparison with more realistic residual interactions,
independently of the SDI strength.

\begin{figure}
\centerline{\psfig{figure=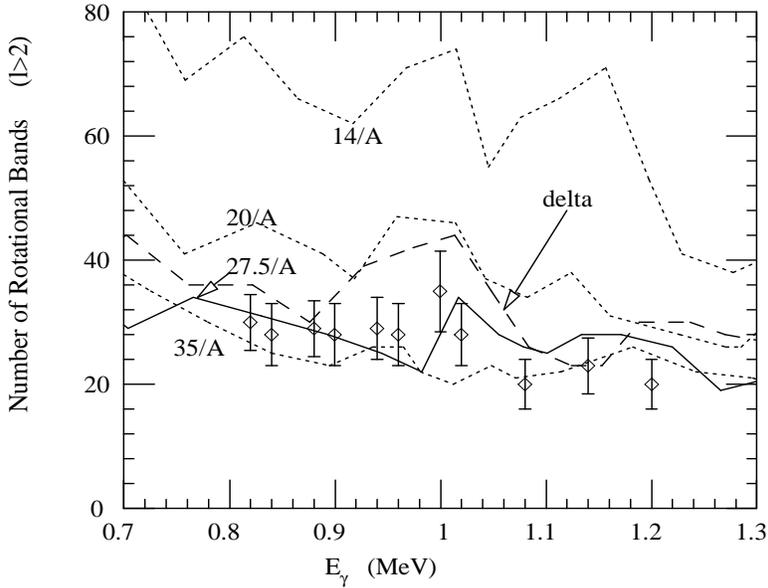,width=10cm,angle=-90}}
\caption{\label{fig15}
Number of rotational bands calculated with
various SDI strengths and with the delta force.
The definition of the number of bands is based on the
criterion that $n_{branch} < 2$ for at least 2 steps.
}
\end{figure}

Figure \ref{fig15} compares results from these forces with respect to
the number of rotational bands taken as an measure of the
onset of rotational damping.
The calculated number of bands is rather stable for 
reasonable values $20/A - 35/A$ MeV of the SDI strength 
(See Appendix B), in 
fairly good agreement with the experimental
effective number of path.
The volume delta force also gives similar results.

In contrast, if the SDI strength is
reduced, e.g., to $V_0 = 14/A$ MeV, the number
of bands increases significantly and overestimates the experimental
data. This is because the configuration mixing effects
is reduced accordingly (e.g. $n_{mixing}$ becomes about factor
four smaller than with $V_0 = 27.5/A$ MeV) and the weakened
 configuration mixing is less capable of causing the rotational
damping. On the other hand,
the number of rotational bands does not decrease very much 
increasing the SDI strength.
This feature is qualitatively
explained  by the role of the
level densities for the onset  of damping. According to
Ref.\cite{Lauritzen}, the onset energy
$U_0$ scales with the
interaction strength $v$ only in powers
$U_0 \sim v^{-2/3}$ while
the number
of levels (or number of rotational bands) below
$U_0$ depends on $U_0$ in an exponential manner
$ exp{\sqrt{2 a U_0}}$ as estimated from
the Fermi gas level density.
It should be remarked, however,  that
the rotational band structure near yrast line is
significantly deviated from the unperturbed rotational bands
(corresponding to the conventional cranking calculations) if the
strength exceeds more than $35/A$ MeV. Such large strength may not
be very realistic for the description of low-lying
rotational bands near the yrast even though the onset of rotational
damping is described fairly well.

\subsection{Role of high multipole components}\label{sec:highmul}

In addition to just the strength of the residual interaction, also the 
spatial properties will affect its ability to generate configuration
mixing. Earlier \cite{Matsuo93} we have shown that a pure pairing plus
quadrupole-quadrupole (P+QQ) interaction 
produces transition strength distributions 
and level spacing distributions which are very different from those 
generated by the SDI. 
The P+QQ force
provides a good description of only the large matrix
elements which lead to collective properties
of low-lying levels.
It is to be noticed however that the low multipole forces
such as P+QQ usually have strong selectivity or selection rules
with respect to the Nilsson asymptotic quantum numbers.
Contrary, the components with high multipolarity may not have such large
matrix elements nor cause the collective effects while
they may have less selectivity.

\begin{figure}[t]
\centerline{\psfig{figure=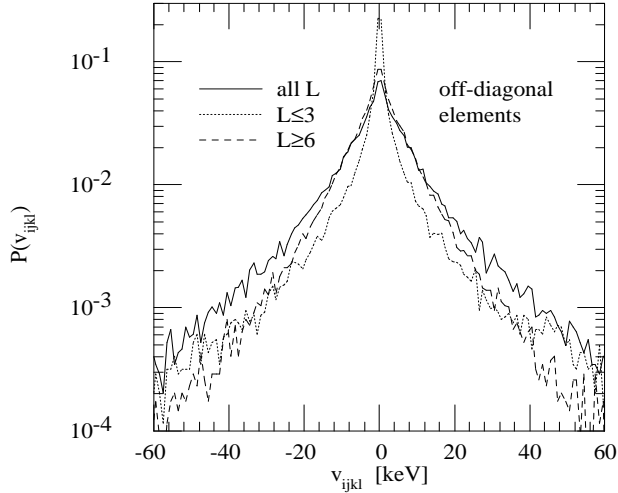,width=8cm,angle=-90}}
\caption{\label{fig16}
Distribution of the off-diagonal two-body
matrix elements $v_{ijkl}$ of SDI with $V_0 = 27.5/A$ MeV,
binned within intervals of width 1 keV.
The angular momentum chosen for the figure is $I=40,41$.
The distribution for the low-multipole part ({$L \le 3$}) and
for the high-multipole part ($L \ge 6$) of the same SDI
are also shown. The latter two have essentially the same r.m.s.
value. 
}
\end{figure}

\begin{table}[b]
\begin{center}
\begin{tabular}{|c | c |c | c|}
\hline
 & r.m.s. (keV)&  &  r.m.s. (keV)  \\
\hline
 all $L$ & 19.8 & & \\
$L \le 2$ & 11.3 & $L \ge 3$ & 16.5 \\
$L \le 3$ & 12.3 & $L \ge 4$ & 15.7 \\
$L \le 4$ & 16.3 & $L \ge 5$ & 13.4 \\
$L \le 5$ & 16.8 & $L \ge 6$ & 12.3 \\
\hline
\end{tabular}
\caption{\label{table1}
The root mean square value $\protect\sqrt{\langle v_{ijkl}^2 \rangle}$
of off-diagonal matrix elements of the residual two-body
force  for various multipole decompositions
of  SDI with $V_0 = 27.5/A$ MeV.
The angular momentum chosen for the table is $I=40,41$.}
\end{center}
\end{table}

Figure \ref{fig16} 
shows the distribution of off-diagonal two-body matrix elements  between
all possible two particle transitions ($ij \leftrightarrow kl$) 
conserving parity and
signature associated with the interaction matrix elements 
$\langle\mu|V_{\rm res}|\mu'\rangle$
between the basis configurations $\{\ket{\mu}\}$. The selectivity of the 
low multipole part $(L \le 3)$ of the SDI manifests itself as a surplus 
of very weak matrix elements, and also of very strong matrix elements.
The restriction to very large multipoles $(L \ge 6)$ is seen to lead
to a more smooth distribution. To investigate how such a selectivity 
affects the configuration mixing, 
we perform several calculations in which only the high or the low
multipole components of the SDI are taken into account.
Low multipole components include $L \leq 2,3,4,5$  in Eq.(\ref{eq:SDI})
with the  SDI strength
$V_0 = 27.5/A$ MeV, and high multipoles with
$L \geq 3,4,5,6$.
Another calculation has also been done using
the P+QQ interaction with the monopole
pairing strength $G_0=20/A$ MeV ,the quadrupole pairing strength
$G_2=2.4 \chi_{self}$  which is estimated from the multipole decomposition
of the delta force \cite{Hamamoto}, and the QQ interaction strength
$\chi_{nn}=\chi_{pp}=0.7\chi_{self}, \chi_{np}= 3.3\chi_{self}$
estimated from the selfconsistency
condition  \cite{Bohr-Mottelson} both for isoscalar and isovector
components. A calculation with the P+QQ interaction with doubled strength is
also performed. To show the average size of the
 the two-body matrix elements, the root mean square (r.m.s.) value
of the off-diagonal elements is listed  in Table \ref{table1}. 
Note that the r.m.s. value associated with the P+QQ 
is  12.7 keV so that  it becomes larger than
the r.m.s. of the full SDI when the force strength
is doubled.

\begin{figure}
\centerline{\psfig{figure=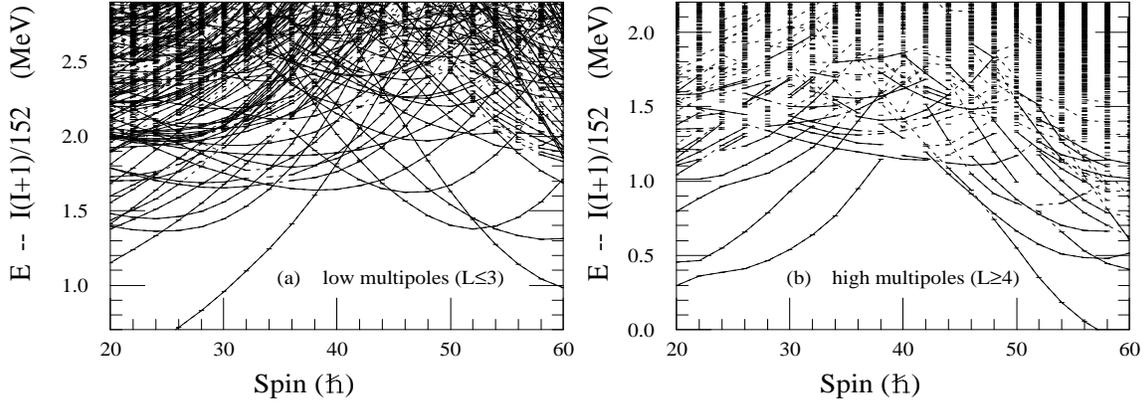,width=15cm,angle=-90}}
\caption{\label{fig17}
The energy levels and rotational band
structure for \Yb $(+,0)$ 
calculated with only (a) the low multipole including
$L \le 3$ terms of SDI in Eq.(\ref{eq:SDI}) and
(b) the high multipoles $L \ge 4$. The strength of SDI
is  $V_0 = 27.5/A$ MeV. See Fig.\ref{fig4} for
definition of solid and dashed lines.
}
\end{figure}

The influence of the different multipolarity 
on the onset of rotational damping
is evident in Fig.\ref{fig17},
in which the rotational band structures are displayed in the same
way as Fig.\ref{fig4} for the calculation with low multipoles
$L \le 3$ of SDI as well as for the one with complementary high multipoles
$L \ge 4$.  With the low multipole part, most of the states
shown in the figure form a large number of
rotational band structures characterized
by  sequences of strong E2 transitions, apparently different
from the calculation Fig.\ref{fig4} with full SDI. On the contrary,
the calculation with the higher multipoles of the SDI gives
essentially the same behavior as the full SDI except
details of individual levels.

\begin{figure}[t]
\centerline{\psfig{figure=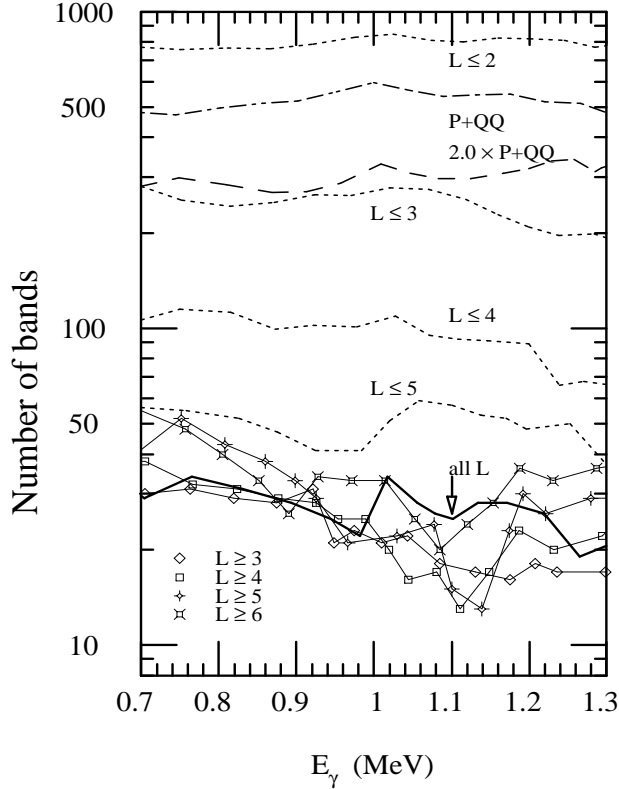,width=8cm,angle=-90}}
\caption{\label{fig18}
The number of rotational bands for various
multipole decompositions of SDI plotted as a function
of average gamma-ray energy (See text for details).
The definition of the number of bands is the  same as in Fig.\ref{fig15}.
The number of bands calculated with the P+QQ force is also
plotted with dotdashed line. The result for the P+QQ with
doubled force strength is plotted with dashed line.
}
\end{figure}

To give a more detailed analysis, the number of bands is calculated
for various choices of multipoles and plotted in Fig.\ref{fig18}.
It is noticeable that the low multipoles
with $L \leq 3$ of SDI as well as the  P+QQ interaction
produce  several hundreds of
rotational bands, which is much larger than the
full multipole SDI and apparently disagree with the
experimental data.
On the other hand, the high multipoles with $L \geq 4,5$
components
give essentially the same number of
rotational bands as the full multipole SDI. These results
point to  that the onset of rotational damping essentially
originates from the high multipole components $L \gesim 4$ of the
two-body residual interaction and the lower multipole components
$L \lesim 3$ contribute little.
It should be noted that the P+QQ with doubled strength
gives still  too large number of bands while it has larger
r.m.s. value of matrix elements than the full multipole SDI.
The low multipoles $L \le 3$ and the high multipoles $L \ge 6$
have the 
same value of r.m.s. while the predicted onset of rotational damping
is completely different.

The origin of this
difference can be traced back to the
distributions of the two-body matrix elements shown
in Fig.\ref{fig16}, where the distribution of off-diagonal two-body matrix
elements $v_{ijkl}$ is  plotted for the SDI with
all the multipoles, with the low multipoles containing  $L \le 3$,
and with the high multipoles
$L \ge 6$
(The same plot for the P+QQ is given in Ref.\cite{Matsuo93}).
The latter two have about the same
r.m.s. value of about 12 keV (cf. Table 1).
All the three interactions have a distribution
which is sharply peaked at $v_{ijkl} = 0$ as compared to the
Gaussian distribution (inverse parabola curve) which is
expected in a random limit. In particular, the distribution
of the low multipole interaction is
significantly concentrated
at small values
$\left| v_{ijkl} \right| \lesim $ a few keV,
indicating  strong selectivity for the matrix elements
of the low multipole interaction
while the high multipole part of the SDI
as well as the full SDI has more uniform
distribution,
On the other hand,
the value of $P(v_{ijkl})$ for the low multipole part
($L \le 3$) of SDI is
factor 2-3 smaller than for the high multipole
part ($L \ge 6$) or the full SDI at
around the average strength ($ v \sim 10-20$ keV) of the
matrix elements.
This implies that the basis $n$p-$n$h configurations $\{ \ket{\mu}\}$
have less chances to interact with each other via
the low-multipole parts of SDI than via the 
high-multipole parts.
In fact, we find that the mixing number
$n_{mixing}$, Eq.(\ref{mixnum}), calculated with the
low multipole part ($L\le 3$) of SDI is significantly
smaller than
those obtained with full SDI or the high multipole part
($L \ge 6)$. For example, the average
value of $n_{mixing}$ calculated at $I^\pi=50^+$
is around 3  even at $U \sim 1.5$ MeV, and many  of
the states around this energy keep a rather pure configuration
with $n_{mixing} \lesim 2$ (cf. $\langle n_{mixing} \rangle
\sim 10$ and $ \sim 8$ for the full SDI and the high multipole
parts $L\ge 6$, respectively). 
Accordingly, one finds that the discrete rotational
bands extend to a much higher excitation energy when 
only the low-multipole parts of SDI are included.

One may consider several reasons for
the distribution of the
two-body matrix elements being much sharper than the Gaussian.
The Gaussian distribution could be obtained if the wave functions of
single-particle orbits had no specific structures
and were randomly distributed, while such a random
situation is not satisfied  for the actual cranked
single-particle orbits under  consideration
since the cranked single-particle orbits keep
some characteristic structures even in the presence
of the deformation and the cranking terms in the mean-field Hamiltonian.
As an example of such structures,  one could  consider
angular momentum  components of single-particle orbits.
Many of the cranked single-particle orbits can be
classified as deformation aligned   or rotation
aligned, depending on whether the angular momentum vector
points to the deformation axis ($z$ axis) or the rotation
axis ($x$ axis), respectively.
The deformation aligned orbitals, which correspond to the almost
straight lines with small slope in Fig.\ref{fig1},
generally carry large
and fairly pure $K$ values (projection of angular momentum along
$z$ axis).
The rotation aligned orbitals  corresponding to
straight lines with large slope are dominated by the
components with large total angular momentum $J$
(e.g. $i_{13/2}$ for neutrons, $h_{11/2},h_{9/2}$,
and $i_{13/2}$ for protons).
These approximate quantum numbers may give rise to
strong selectivity in the two-body matrix elements.
This effect is expected to be stronger for low multipole
interactions than for those containing high multipole contributions.
If the angular quantum numbers
$K$ and $J$ of the single-particle orbits are good quantum numbers
and the
two-body interaction with multipolarity $L$ is considered,
there is a selection rule for the $K$ and $J$ of the
interacting two particles such as $|K_1-K_2|\le L$ and
$|J_1\pm J_ 2| \le L$.
This selection becomes effective for $L$ which
is smaller than the average $J$ of
single-particle orbits near the Fermi surface.

\section{Conclusions}\label{sec:concl}

We have presented results of shell model
calculations for rapidly rotating   warm  nuclei.
The model describes microscopically
levels and stretched E2 transitions up to a few MeV above yrast
at high spin region $ I \gesim 20$. From the numerical calculation for
\Yb, it is predicted that
the rotational damping sets in  around
$U \gesim 0.8$ MeV above the yrast line.
The levels near the yrast line form rotational band structures
and the number of the
rotational bands is calculated to be around 30 at a given rotational
frequency, agreeing with the experimental data extracted from the
fluctuation analysis of  $E_\gamma \times E_\gamma$
spectra.
 
The model  predicts some novel features of the
rotational damping. The onset of rotational damping
takes place quite gradually as a function of internal excitation
energy and it shows large fluctuations depending on individual
states. The transition region extends from 0.8 MeV to 2 MeV above
the yrast line. Even in the region  $U \sim
0.8 - 1.5$ MeV, there remain  scars of discrete rotational band structures,
which are characterized by short sequences of  strong E2
transitions. In keeping with these results, the corresponding strength function
associated with two-fold E2 rotational transitions is expected to display
a two-component profile with a narrow component generated
by strong and correlated transitions surviving in the region of the
rotational damping  and a wide component whose width is
related to the rotational damping width.
The origin of rotational damping can be traced back to
the high-multipole components ($L \gesim 4$) of
the residual two-body force acting among the unperturbed cranked shell model
states.
 
The overall aim of the present work has been to extend the cranked
mean field to finite temperatures, including an effective two-body
force which is capable of coupling many-particle many-hole
excitations built upon the cranked mean field.
The main advantage of the
model is the inclusion of all excitations of the independent particle
motion up to around 2 MeV above yrast line,
and this implies a realistic behavior of the level density.
As a drawback, the truncation of the 
shell model space adopted in the present
paper does not allow for
surface vibrations and pair-correlated states to be generated.
However, since the effective two-body force
applied includes terms which generate such correlations, it would
be an interesting future subject to extend the shell model space
and to study whether the shell model approach is
able to describe also the correlated states at high spins.

\vspace{10mm}
\section*{Acknowledgment}
We deeply acknowledge  fruitful and stimulating discussions
with B. Herskind, A. Bracco, and S. Leoni.
We also thank S. Frauendorf and Y.R. Shimizu for providing us the
cranked Nilsson and the liquid drop codes, respectively.
One of the author, M.M., acknowledges the Danish Research Council
as well as the INFN
for support of his stay at Niels Bohr Institute and INFN Sez. Milano
where
a part of the research reported here was carried out.

\vspace{10mm}
\section*{Appendix A. Diabatic cranked Nilsson single-particle basis}

\setcounter{equation}{0}
\renewcommand{\theequation}{A.\arabic{equation}}

The adiabatic basis or the eigen solutions of the cranked Nilsson
Hamiltonian sometimes show abrupt changes in the routhian energy
$\{e'_i(\omega)\}$ and wave functions
$\{\psi_i(\omega)\}$ when the rotational frequency is varied.
This is caused by the ``repulsive interaction'' associated with
crossings among two orbits with the same quantum numbers.
The diabatic single-particle basis can be constructed by
removing the unwanted repulsion at each of the crossings.

First, we need to specify the unwanted crossings which cause
the abrupt changes in the adiabatic basis. In order to
measure the abruptness, it is necessary to give a relevant
scale for the changes in the rotational frequency. Since
the rotational band structure associated with the collective
rotation is concerned, we consider the change in the
frequency corresponding to
the angular momentum change $\Delta I = 2$ of two unit.
Namely, we examine the change in $\{e'_i(\omega)\}$ and
$\{\psi_i(\omega)\}$  against the variation from  $\omegaI$ to
$\omegaII$, where $\omegaI$ is the rotational frequency corresponding
to a given angular momentum $I$ (See, sect.\ref{sec:base}). 
The relation between $\omega$ and $I$
approximately follows $I \sim J \omega$,  $J$ being
the moment of inertia. 

Now the unwanted crossings are specified by the following three
criteria. i) The single-particle wave functions at $\omegaI$ and
$\omegaII$ should not be very different for the diabatic
basis. Given an orbit $i$ at $\omegaI$, an overlap condition
\beq
\left|\langle \psi_i(\omegaI)| \psi_j(\omegaII)\rangle\right| > \Omega_{thr}
\eeq
is checked for orbits $j$ at $\omegaII$. If there is no
orbits satisfying this condition, we consider the orbit $i$
as interacting at $\omegaI$. ii) The single-particle contribution
to the dynamic moment of inertia
$j_i^{(2)} =-d^2e'_i/d^2\omega=dj_{x,i}/d\omega$
is calculated at $\omegaI$. If its absolute
magnitude exceeds a threshold $j_{thr}$, the orbit
$i$ is considered as interacting at $\omegaI$ since
the moment of inertia for the configurations involving this orbit
deviates significantly from the other configurations.
iii)
Under the assumption that only two diabatic levels interact
simultaneously with a constant interaction, all the
features of the level repulsion are specified
by the interaction strength
$v_{int}$, the crossing frequency $\omega_{crs}$ and
another quantity $\delta$  which
is the relative energy shift of diabatic levels under the
change $\omegaI \rightarrow \omegaII$ in the rotational frequency.
A measure of the abruptness of the crossing is given by
a dimensionless parameter $v_{int}/\delta$ \cite{Bengtsson}.
For small $v_{int}/\delta$, below a threshold $e_{thr}$,
the adiabatic basis
shows an abrupt change near the crossing point. Here
the interaction strength $v_{int}$ and the crossing
frequency $\omega_{crs}$ are  extracted by searching the frequency where
the distance
$|e'_1(\omega) - e'_2(\omega)|$
between the interacting orbit pair is minimum.
The interaction strength $v_{int}$ is a half of the minimum distance.
The parameter $\delta$ is expressed in terms of  the
diabatic energies of the interacting orbits, i.e.,
\beq
\delta = 1/2\{e'_{{\rm dia},i}(\omegaI) - e'_{{\rm dia},i+1}(\omegaI)
 - (e'_{{\rm dia},i}(\omegaII) - e'_{{\rm dia},i+1}(\omegaII))\}
\eeq
where the expression for the diabatic energy is given by (A.5).
The conditions i) and iii) are essentially the same as those
introduced by Bengtsson \cite{Bengtsson}.

The three criteria discussed above represent similar conditions
about the abruptness, and
are not mutually independent.
In practice, we first pick up candidates
with use of the first two criteria, and then examine  them
in terms of the third criterion in order to finally determine
the pairs of
orbits for which the diabatic basis should be constructed.
A reasonable choice of the thresholds is found to be
$\Omega_{thr}=0.933   , j_{thr}=20$ MeV${}^{-1}$ and $e_{thr}=1.8$.

Once the unwanted crossings are specified, we then remove the
interaction among the two crossing orbits. In doing this,
we assume that the crossings are isolated and that each crossing
is represented by a two-level model with a constant interaction
strength \cite{Takami}.
For the two-level model, the relation between the adiabatic
and the diabatic basis is analytically expressed. The routhian
and the wave function of the diabatic basis are expressed
as
\beq
e'_{{\rm dia},1 \ {\rm or}\  2}=
\pm \sqrt{{(e'_1 - e'_2 )/2}^2 - v_{int}^2}
 + (e'_1 + e'_2)/2
\eeq
\beqa
\psi_{{\rm dia},1} & = \cos(\theta)\psi_1 - \sin(\theta) \psi_2 \\
\psi_{{\rm dia},2} & = \sin(\theta)\psi_1 + \cos(\theta) \psi_2
\eeqa
with
\beq
\cos(\theta) = \sqrt{\left(1 + {e'_{{\rm dia},1}-e'_{{\rm dia},2}
\over e'_1-e'_2}\right)/2}
\eeq
in terms of the 
eigensolutions of the cranked Nilsson single-particle
Hamiltonian, which form the adiabatic base.
This prescription
is applied to the interacting orbit pairs
within a frequency interval between $\omega_{min}$ and $\omega_{max}$,
at which the distance between the interacting adiabatic orbits
becomes 10 times the interaction strength $v_{int}$.
There is only small
difference between the diabatic and adiabatic basis outside of this
frequency interval, where the latter should be used.

\section*{Appendix B.  Surface Delta Interaction}

\setcounter{equation}{0}
\renewcommand{\theequation}{B.\arabic{equation}}

The cranked Nilsson single-particle basis is expanded in the
harmonic oscillator basis
$\varphi_{nljm}(r_t \theta_t \phi_t)$
$= R_{nl}(r_t)\varphi^{angle}_{ljm}(\theta_t \phi_t)$
in the single-stretched polar coordinate
system $\{r_t \theta_t \phi_t\}$ where the
single-stretch means
the scaling ${x_i = \sqrt{\omega_0/\omega_i}x'_i}$ with
the anisotropic oscillator frequencies of the Nilsson
potential. Correspondingly, the SDI
can be expressed in terms of the stretched coordinates
\beqa
v(1,2) & = & - V'\delta(\vec{x}'_1 - \vec{x}'_2)\delta(r_t - R) \\
       & = & - V'{\delta(r_{t,1}-R) \over r_{t,1}}
        {\delta(r_{t,2}-R) \over r_{t,2}}\delta(\Omega_{t,12})
\eeqa
where $\delta(\Omega_{t,12})$ is the delta function for the
angle variables.
As it is usually prescribed, the radial matrix elements of the SDI
interaction is replaced by a constant. Then the SDI interaction
acting on the angular and spin
wave function $\varphi^{angle}_{ljm}$ is given
\beq \label{eq:SDI}
v(1,2)^{angle} =  - 4\pi V_0 \sum_{LM}Y^{*}_{LM}(\theta_{t,1} \phi_{t,1})
Y_{LM}(\theta_{t,2} \phi_{t,2})
\eeq
The evaluation of the angular matrix elements of the
spherical harmonics is straightforward  \cite{Brussard}.
The matrix elements in
the cranked Nilsson orbits are then calculated directly.

The interaction strength $V_0$
is directly related to the strength $G_0$ of the standard monopole
pairing force \cite{Mozkowski}, for which the value around
$20/A - 30/A$ MeV is often used. The requirement of the selfconsistency
between induced mean-field and density gives an
estimate $V_0 = 20/A$ MeV \cite{Faessler}.
{}From fits in light nuclei
($A \lesim 70$), it is estimated that $V_0 \sim 25/A$ MeV \cite{Brussard}.
The analysis of 
the low-lying collective vibrations and the pairing
in rare-earth deformed nuclei gives 
$V_0 = 27.5\pm6.5/A$ MeV \cite{Faessler}.
Thus, we  adopt $V_0 = 27.5/A$ MeV as a representative
strength of SDI.

\end{document}